\documentclass[useAMS,usenatbib]{mn2e}
\usepackage{epsfig}
\usepackage{color}
\usepackage{deluxetable}
\usepackage{rotating}
\usepackage{float}
\usepackage{amssymb}
\usepackage{graphicx}
\usepackage{subfigure}
\usepackage{journals}
\usepackage{psfrag}
\usepackage{amsmath}
\usepackage{amssymb}
\usepackage{lscape}
\usepackage{IEEEtrantools}
\usepackage[colorlinks=true,citecolor=blue]{hyperref}
\def\reff@jnl#1{{\rm#1\/}}
\def\aj{\reff@jnl{AJ}}                 
\def\araa{\reff@jnl{ARA\&A}}           
\def\apj{\reff@jnl{ApJ}}               
\def\apjl{\reff@jnl{ApJ}}              
\def\apjs{\reff@jnl{ApJS}}             
\def\ao{\reff@jnl{Appl.Optics}}        
\def\apss{\reff@jnl{Ap\&SS}}           
\def\aap{\reff@jnl{A\&A}}              
\def\aapr{\reff@jnl{A\&A~Rev.}}        
\def\aaps{\reff@jnl{A\&AS}}            
\def\azh{\reff@jnl{AZh}}               
\def\baas{\reff@jnl{BAAS}}             
\def\jrasc{\reff@jnl{JRASC}}           
\def\memras{\reff@jnl{MmRAS}}          
\def\mnras{\reff@jnl{MNRAS}}           
\def\pra{\reff@jnl{Phys.Rev.A}}        
\def\prb{\reff@jnl{Phys.Rev.B}}        
\def\prc{\reff@jnl{Phys.Rev.C}}        
\def\prd{\reff@jnl{Phys.Rev.D}}        
\def\prl{\reff@jnl{Phys.Rev.Lett}}     
\def\pasp{\reff@jnl{PASP}}             
\def\pasj{\reff@jnl{PASJ}}             
\def\qjras{\reff@jnl{QJRAS}}           
\def\skytel{\reff@jnl{S\&T}}           
\def\solphys{\reff@jnl{Solar~Phys.}}   
\def\sovast{\reff@jnl{Soviet~Ast.}}    
\def\ssr{\reff@jnl{Space~Sci.Rev.}}    
\def\zap{\reff@jnl{ZAp}}               
\def\nat{\reff@jnl{Nature}}            
\def\Ha{H$\alpha$~}
\def\HI{H{\small{I}}~}
\title[Blue compact dwarf galaxy Mrk~22]{ Tidal interaction, star formation and chemical evolution in blue compact dwarf galaxy Mrk~22}
\author[A. Paswan, A. Omar and S. Jaiswal]{A.~Paswan \thanks{E-mail:
p.abhishek@aries.res.in}, A. Omar \thanks{E-mail:
aomar@aries.res.in} and S. Jaiswal\\ Aryabhatta Research Institute of Observational Sciences, Manora Peak, Nainital 263002, India\\ Pt. Ravishankar Shukla University, Raipur, 492010, India}
      
\begin{document}

\date{Accepted ------------, Received ------------; in original form ------------}
\pagerange{\pageref{firstpage}--\pageref{lastpage}} \pubyear{}
\maketitle
\label{firstpage}

\begin{abstract}

The optical spectroscopic and radio interferometric \HI 21 cm-line observations of the blue compact dwarf galaxy Mrk~22 are presented. The Wolf-Rayet (WR) emission line features corresponding to high ionization lines of He{\small{II}}~$\lambda$4686 and C{\small{IV}}~$\lambda$5808 from young massive stars are detected. The ages of two prominent star forming regions in the galaxy are estimated as $\sim$10 Myr and $\sim$ 4 Myr. The galaxy has non-thermal radio deficiency, which also indicates a young star-burst and lack of supernovae events from the current star formation activities, consistent with the detection of WR emission lines features. A significant N/O enrichment is seen in the fainter star forming region. The gas-phase metallicities [12 + log(O/H)] for the bright and faint regions are estimated as 7.98$\pm$0.07 and 7.46$\pm$0.09 respectively. The galaxy has a large diffuse \HI envelop. The \HI images reveal disturbed gas kinematics and \HI clouds outside the optical extent of the galaxy, indicating recent tidal interaction or merger in the system. The results strongly indicate that Mrk~22 is undergoing a chemical and morphological evolution due to ongoing star formation, most likely triggered by a merger. 

\end{abstract}

\begin{keywords}
galaxies: starburst - galaxies: interaction - galaxies: abundances - galaxies: dwarf - stars: Wolf-Rayet 
\end{keywords}


\section{Introduction}

The blue compact dwarf (BCD) galaxies show a blue appearance with compact (\textless 1 kpc) star-bursting regions \citep{1965ApJ...142.1293Z,1981ApJ...247..823T}. These galaxies have low stellar-mass (M$_{*}$ $\lesssim$ 10$^{10}$ M$_{\odot}$), low luminosity (M$_{B}$ $\gtrsim$ -18 mag) and high gas content \citep[M$_{HI}$ $\gtrsim$ 10$^{8}$ M$_{\odot}$;][]{2013ApJ...764...44Z}. BCD galaxies have low median oxygen abundance [12 + log(O/H)] $\sim$ 8.0 with a range between 7.0 and 8.4 \citep{1972ApJ...173...25S,1999ApJ...527..757I,2000A&ARv..10....1K,2004ApJS..153..429K,2008A&A...491..113P}. The typical morphology of BCDs consists of a few distinct knots of star formation. The optical spectrums of BCDs are dominated by strong emission lines attributable to the ongoing star formation. The very young ($\lesssim$ 10 Myr) stellar masses dominated by O/B type stars in the star-bursting regions give blue colours to these galaxies. A majority of BCDs also have an underlying old (\textless 10 Gyr) stellar component of dominant stellar masses, implying that BCDs are not young systems \citep{2011AJ....141...68Z}. The average contribution of young stellar components to the optical emission in a BCD is $\sim$ 50 per cent, and sometimes up to 90 per cent in extremely metal poor BCDs \citep{1999AGM....15.P102N,2001ApJS..136..393C,2009A&A...501...75A}. The starbursts in BCD galaxies often do not last longer than about 10 Myr and are separated by relatively long phases of quiescence \citep{1991ApJ...370...25T,1995A&A...303...41K,1999A&A...349..765M,2000ApJ...539..641T}. The optical colours of several isolated dwarf irregulars suggest on average a low star formation rate in a quasi-continuous manner with a gas depletion time longer than Hubble time \citep{2001AJ....121.2003V}. 

The star formation in BCDs is fed by a relatively large amount of gas \citep{1981ApJ...247..823T,1992MNRAS.258..334S,1998AJ....116.1186V}. The metal deficient BCD galaxies, which are embedded in large presumably primordial \HI cloud may provide strong clues about the trigger of star formation in young galaxies \citep{1995ApJ...445..108T,1999ApJ...527..757I,2010MNRAS.403..295E}. The mergers and interactions seem to play an important role in triggering star formation in dwarf galaxies \citep{2013seg..book..555S,2014Natur.507..335A,2014ApJ...794..115D}. Various studies indicate that the most probable cause of recent starbursts in BCDs is tidal interactions with nearby dwarf or \HI cloud without a clear optical counterpart \citep{2001A&A...371..806N,2001A&A...374..800O,2001ApSSS.277..445P,2008MNRAS.388L..10B,
2008MNRAS.391..881E,2010MNRAS.403..295E,2016MNRAS.462...92J}. These interacting dwarf galaxies and \HI clouds may have very low optical luminosity and therefore may be missing in the present generation optical surveys. The gas accretion from inter galactic medium (IGM) is also considered a probable cause of starbursts, primarily in extremely metal poor galaxies \citep{1999ApJ...514...77N,2006MNRAS.370.1445D,2009ApJ...703..785D,2012RAA....12..917S}. It is also believed that gas-rich dwarf irregular galaxies can be formed by material ejected from the disks of their parent massive galaxies to the IGM by tidal forces \citep{2000ApJ...543..149O}. In some cases, such star forming BCDs are found at the end of long stellar tails in interacting systems \citep{2004A&A...427..803D,2007A&A...475..187D}. 

The interaction features are detected easily in the \HI 21 cm-line in comparison to the optical images \citep{2010A&A...521A..63L,2012MNRAS.419.1051L,2014A&A...566A..71L,2015ApJ...815L..17M}. Tidal interactions can also be inferred  by identifying multiple nuclei and arcs like structures in the optical bands \citep{1993ApJS...85...27M}. Many galaxies which are considered optically isolated often show prominent signs of interaction in the \HI images. The kinematical disturbances in the \HI disk also reveal very crucial information on the progressive stage of tidal interaction. The \HI imaging is therefore fundamentally important for understanding starburst trigger in galaxies, particularly in optically isolated star-bursting BCD galaxies. The star forming dwarf galaxies (SFDGs) are very crucial objects in understanding cosmic structure formation and evolution as more than 70\% of all galaxies in the local universe are believed to be SFDGs \citep{2004AJ....127.2031K}. The hierarchical formation models of galaxies viz., \citet{1997MNRAS.286..795K} support formation of galaxies via merger. In this context, the BCDs are important objects to study causal phenomenon between the star formation and mergers and/or interactions \citep{2004AJ....127..264B,2008MNRAS.388L..10B,2012MNRAS.419.1051L,2014AJ....148..130A}. It is also believed that the Lyman continuum photon-leaks from extreme SFDGs may be the main source of ionization of the IGM \citep{2009ApJ...693..984W,2013A&A...553A.106L,2013MNRAS.428L...1M,2014Sci...346..216B,2016Natur.529..178I}. 

The chemical abundance patterns in BCD galaxies are quite complex and depend on the current star formation rate (SFR) as well as the history of star formation \citep{2006MNRAS.372.1069M}. The spatially resolved studies of several BCDs indicate near coeval star formation events in multiple H{\small{II}}~regions within them without any extraordinary gradient in abundances such as N/O ratio \citep{2013AdAst2013E..20L}. This suggests an efficient transport either by expanding starburst-driven supershells or gas infall from the halo. The spatially resolved abundances in BCD can provide important clue about the evolution of galaxies. The highly debated issues for understanding evolution of BCDs are gas accretion and trigger of star formation, chemical enrichment through stellar feedback, and subsequent processes leading to quiescence \citep{1985ApJ...299..881T,1988MNRAS.233..553D,1998MNRAS.299..249S,2000MNRAS.313..291F,2002A&A...389..367T,
2006A&A...445..875R,2008MNRAS.388L..10B,2009A&A...508..615L,2013AdAst2013E..20L,2016MNRAS.462...92J}.

In this paper, we present a detailed analysis of Mrk~22 based on optical spectroscopic observations, HI 21 cm-line observations, and archival data in radio and optical bands. Mrk~22 is a metal poor (12 + log(O/H) $\sim$ 8), gas-rich (M$_{HI}$ $\sim$10$^{8}$ M$_{\odot}$), Wolf-Rayet (WR) BCD galaxy \citep{1994ApJ...435..647I,1999ApJ...527..757I,1999AJ....117.2789H,2002AJ....124..862H,2010ApJ...710..663Z,
2011AJ....141...68Z,2013ApJ...764...44Z}. A broad nebular emission line in the galaxy spectrum was detected, but the spectrum appeared noisy to detect C{\small{IV}}~$\lambda$5808 broad emission feature \citep{1994ApJ...435..647I}. Mrk 22 has an irregular outer halo and an off-centre nucleus with an iI morphology \citep{1986sfdg.conf...73L}. The mean stellar ages of young and old components were estimated as $\sim$ 4.7 Myr and $\sim$ 5.6 Gyr respectively \citep{2011AJ....141...68Z}. Mrk~22 shows a double nuclei nature in its B-band optical image with a bright nucleus and a faint diffuse region at about 0.5 kpc separation \citep{1993ApJS...85...27M}. It is considered as a probable merger system \citep{2014ApJ...784...16M}. The H${\alpha}$ morphology of Mrk~22 with estimates for SFRs were presented in \citet{2016MNRAS.462...92J}. The blue colour in the diffuse region is attributed to the ongoing star formation. The main properties of Mrk 22 are listed in Table 1. This paper presents spatially resolved abundance analysis, radio continuum detection, and \HI images of Mrk~22. Some important results also highlight possible scenarios for evolution of BCD galaxies.

\begin{table}\label{tab:01}
\centering
\caption{General properties of Mrk~22}
\begin{tabular}{c|c}
\hline
Paramater & Mrk~22\\
\hline
RA [J2000]$^{a}$ & 09$^{h}$ 49$^{m}$ 30.3$^{s}$\\
Dec [J2000]$^{a}$ & +55$^{d}$ 34$^{m}$ 47$^{s}$\\
Morphological type$^{b}$ & iI\\
Distance [Mpc]$^{a}$ & 23.4$\pm$1.7\\
V$_{helio}$ [km s$^{-1}$]$^{a}$ & 1551$\pm$12\\
E(B--V)$_{Galactic}$ [mag]$^{a}$ & 0.06$\pm$0.04\\
M$_{B}$ [mag]$^{b}$ & -15.82\\
m$_{R}$ [mag]$^{b}$ & 15.80 $\pm$ 10.10\\
(B--R) [mag]$^{b}$ & 0.19 $\pm$ 0.11\\
Log M$_{gas}$ [M$_{\odot}$]$^{c}$ & 8.57 $\pm$ 0.18\\
Log M$_{\star}$ [M$_{\odot}$]$^{c}$ & 7.97\\
Gas fraction$^{c}$ & 0.80 $\pm$ 0.87\\
S$_{1.4 GHz}$ [mJy]$^{d}$ & \textless 2.2\\
SFR$_{1.4 GHz}$ [M$_{\odot}$ yr$^{-1}$]$^{d}$ & \textless 0.24\\
Log L$_{1.4 GHz}$ [L$_{\odot}$]$^{d}$ & \textless 2.72\\ 
Log L$_{FIR}$ [L$_{\odot}$]$^{e}$ & \textless 8.29\\  
\hline
\end{tabular}\\
\begin{flushleft}
$^{a}$Parameters taken from NASA/Extragalactic Database (NED),
$^{b}$\citet{2003ApJS..147...29G},
$^{c}$\citet{2013ApJ...764...44Z},
$^{d}$\citet{2002AJ....124..862H},
$^{e}$\citet{1991AJ....101.2034M}. 
\end{flushleft}
\end{table}


\section{Observations and results}

\subsection{Optical}

The Faint Object Spectrograph and Camera (FOSC) on the 2-m Himalayan Chandra Telescope (HCT) of the Indian Astronomical Observatory (IAO), Hanle, India was used to carry out optical spectroscopic observations. The HCT FOSC is equipped with a 2k $\times$ 4k SITe CCD chip, which uses the central 2k $\times$ 2k region with a plate scale of 0.296\arcsec~pixel$^{-1}$. The gain and readout noise of the CCD camera are 1.22 $e^{-}$ per ADU and 4.87 $e^{-}$ respectively. The spectroscopic observations of Mrk~22 were obtained with a slit of aperture 1.92\arcsec $\times$ 11\arcmin~and a grism providing a spectral resolution of $\sim$ 1330. The spectrum covers the wavelength range from $\sim$ 3500 \AA~to $\sim$ 7500 \AA~ with a dispersion of $\sim$ 1.5 \AA~ pixel$^{-1}$ and an effective spectral resolution of $\sim$ 11 \AA~. The slit position was located at the position angle (P.A.) of $\sim$ 45$^{\mathrm 0}$ in order to cover the full extent of the star forming regions along the major axis of the galaxy. The Fe-Ar lamp exposures  were used for the wavelength calibration of the spectrum. The absolute flux calibration was achieved by observing spectrophotometric standard star HR 7596 selected from \citet{1990AJ.....99.1621O}. The observations were carried out on May 11, 2013 by acquiring two exposures of the target source with an integration time of nearly 1800 sec and 3150 sec on an average airmass of 1.2. The spectroscopic data reduction was performed using the standard procedures in Image Reduction and Analysis Facility {\small (IRAF)}. Bias-subtraction and flat-fielding were applied on each frame. Cosmic ray removal was done using Laplacian kernel detection algorithm \citep{2001PASP..113.1420V} in each frame. Extraction of one-dimensional spectra based on optimal extraction algorithm by \citet{1986PASP...98..609H} was carried out. The spectrum was transferred to the rest-frame of Mrk~22. The line fluxes were obtained using {\small DEBLEND} task of the {\small IRAF}.

\begin{figure*}
\centering
\includegraphics[width=8.0cm]{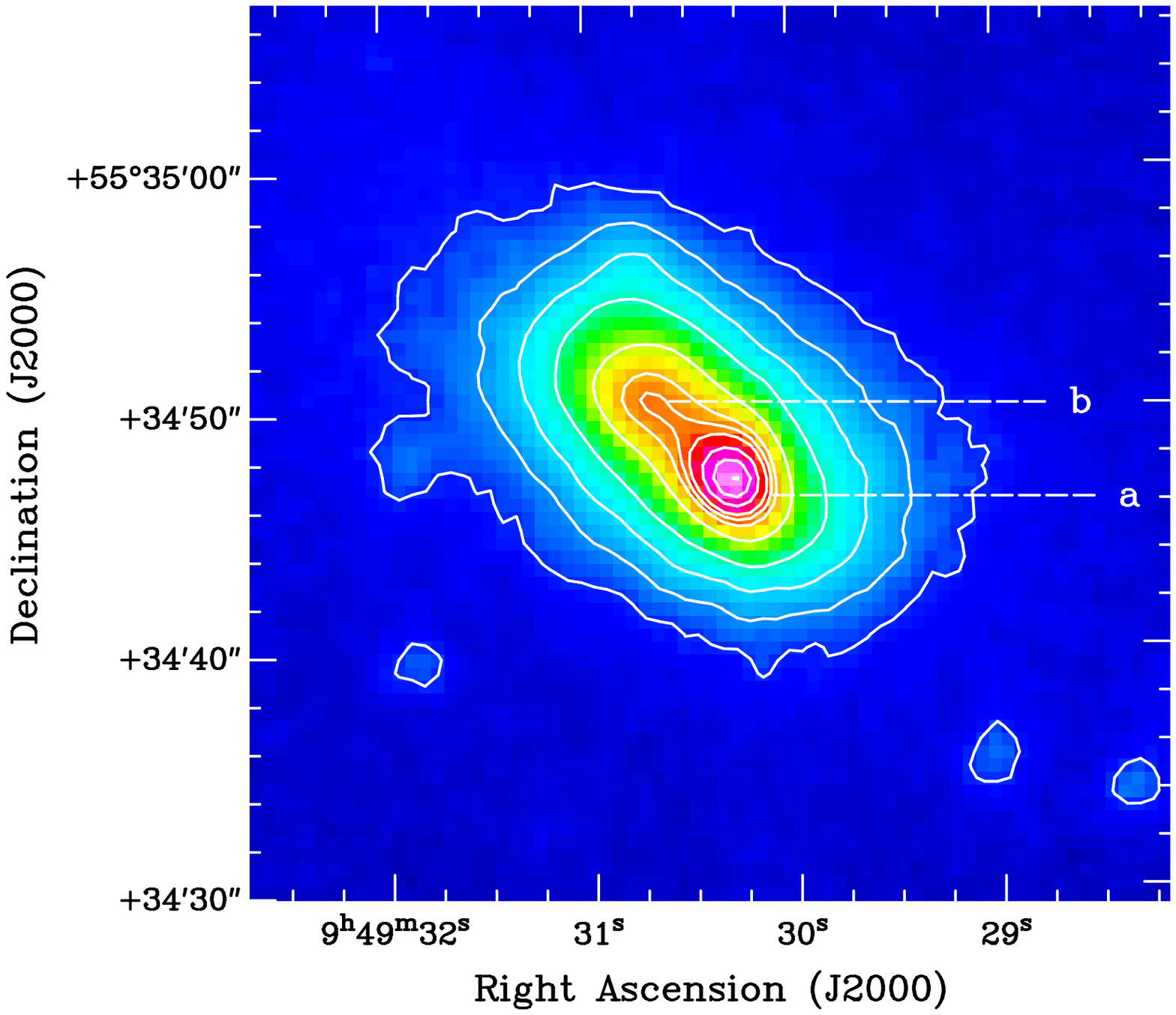}
\includegraphics[width=8.71cm,trim={0 5.3cm 0 3.65cm},clip]{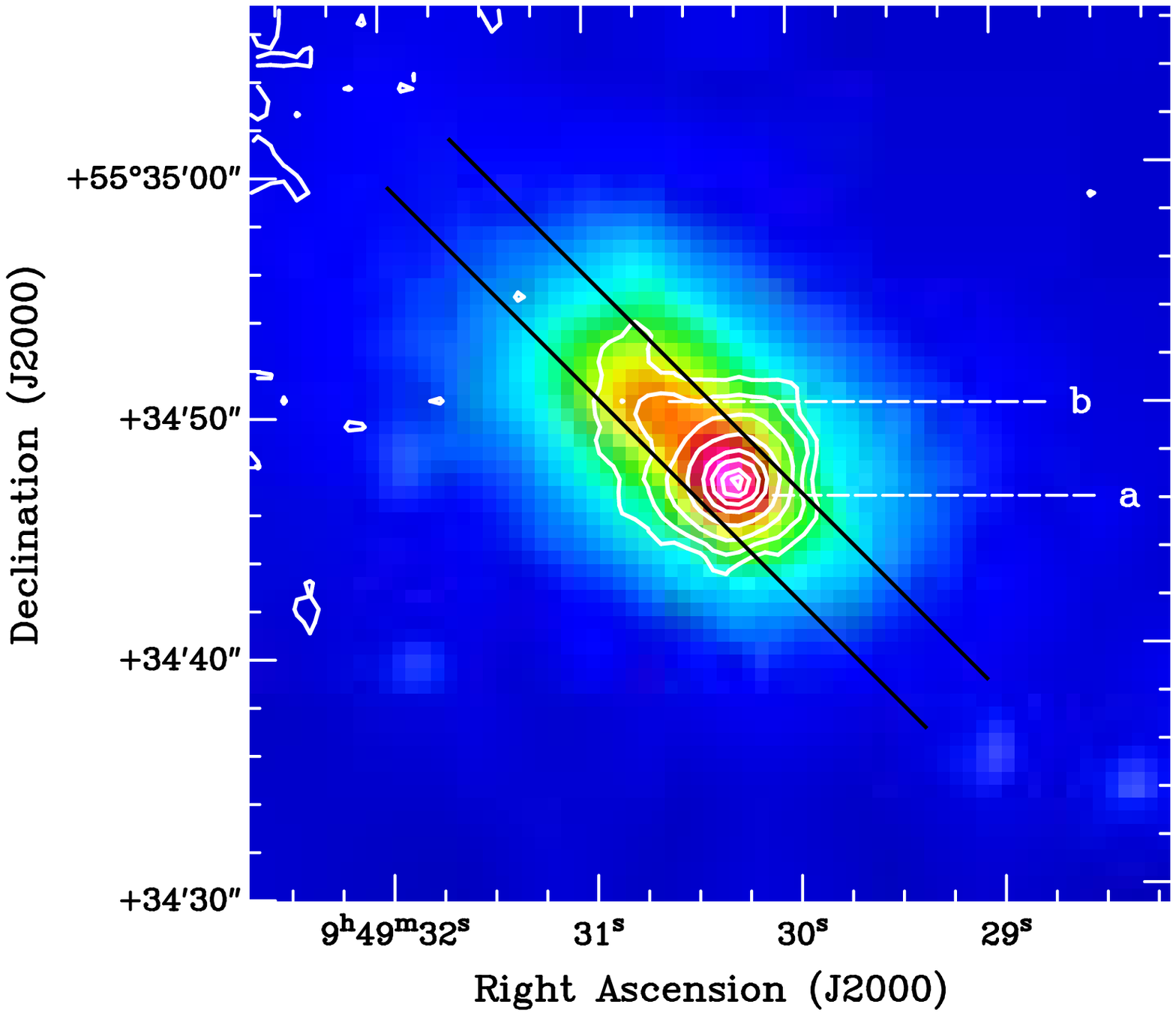}
\caption{The SDSS r-band contour image (left) and the continuum subtracted H$\alpha$ contour image (right) taken with the 1.3-m Devasthal Fast Optical Telescope, overlaid upon the SDSS r-band image of Mrk~22.}
\label{fig:01}
\end{figure*}

\begin{figure*}
\centering
\includegraphics[width=6.5cm, angle=270]{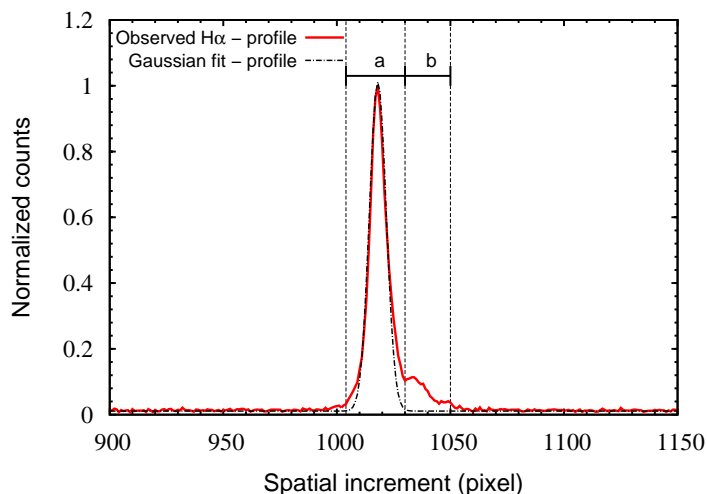}
\caption{The spatial H${\alpha}$ profile along the length of the slit. Two apertures are labeled as 'a` and 'b` over which separate spectrums were extracted. A Gaussian fit to the bright peak is also plotted.}
\label{fig:02}
\end{figure*}

\begin{figure*}
\centering
\includegraphics[width=5.5cm,height=17.0cm,angle=270]{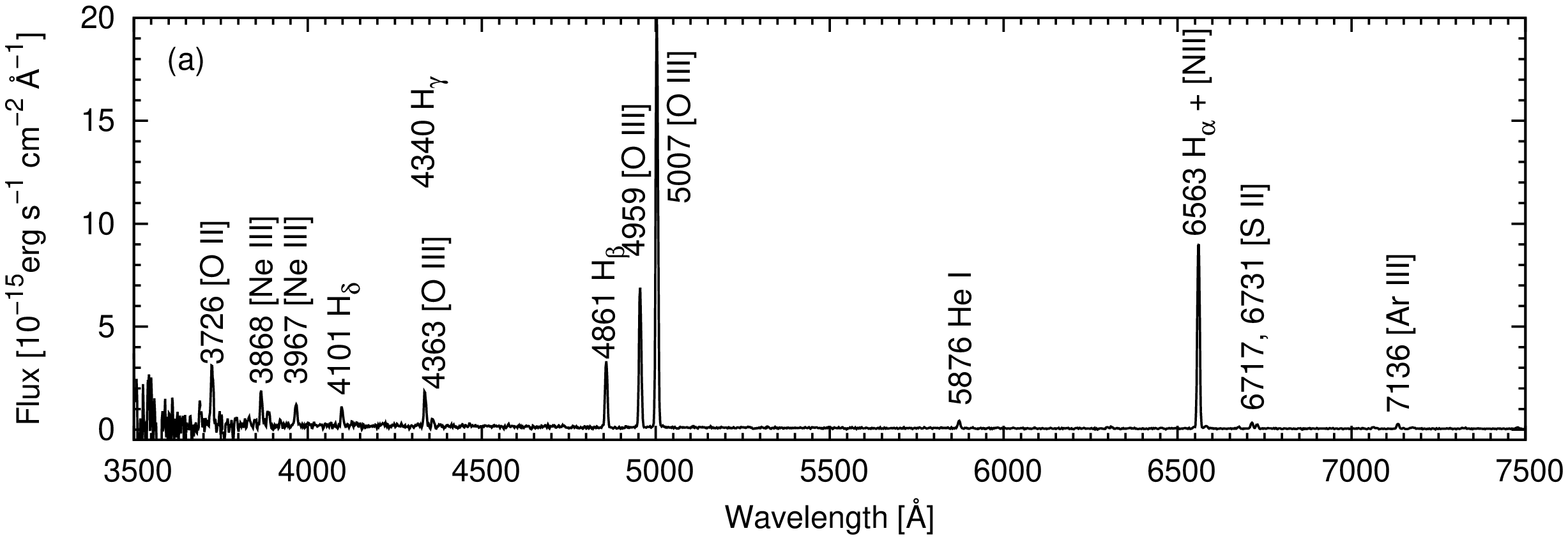}
\includegraphics[width=5.5cm,height=17.0cm,angle=270]{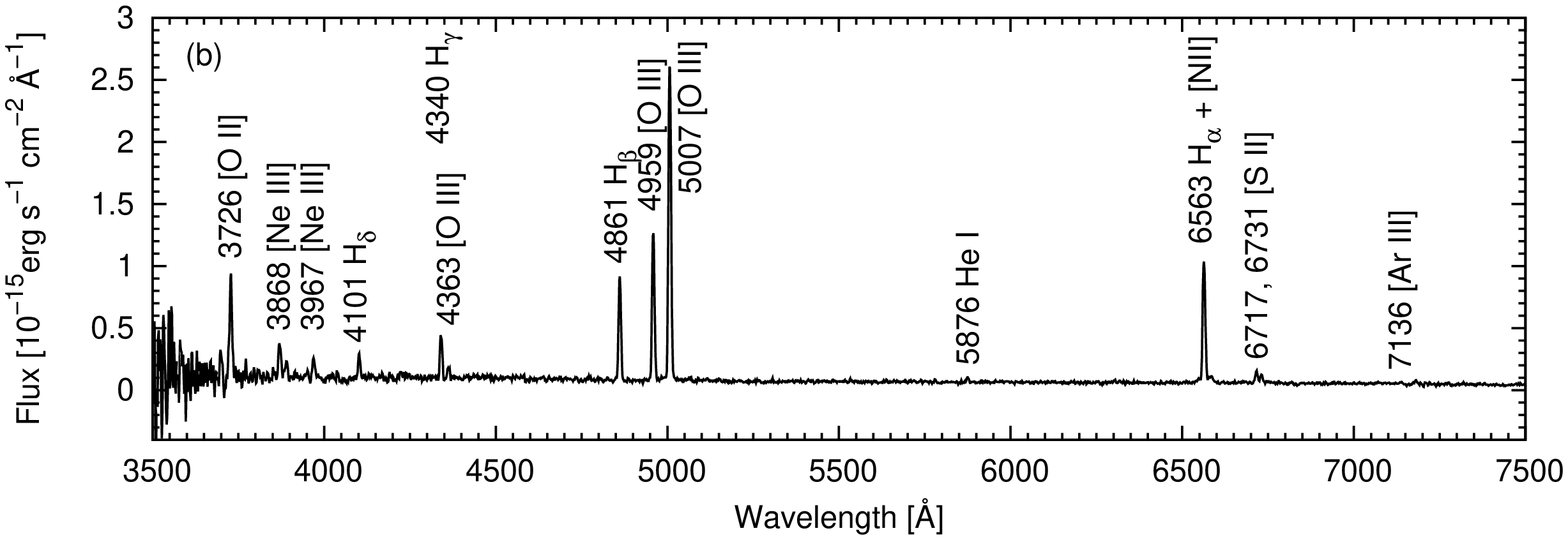}
\caption{The rest-frame de-reddened optical spectrum of (a) bright and (b) faint star-forming region of Mrk~22, taken with the 2-m Himalayan Chandra Telescope.}
\label{fig:03}
\end{figure*}

The star forming regions in Mrk~22 were identified from the continuum subtracted H${\alpha}$ image taken with the 1.3-m Devasthal Fast Optical Telescope (DFOT), ARIES, Nainital, India \citep{2016MNRAS.462...92J}. The typical seeing full width at half-maximum (FWHM) of the H$\alpha$ observation was $\sim$ 2.1\arcsec. The contour maps of the SDSS (Sloan Digital Sky Survey) r-band and the continuum subtracted H${\alpha}$ band images overlaid upon the SDSS r-band image are shown in Fig.~\ref{fig:01}. The SDSS r-band contour map clearly shows two prominent emission regions marked as knot `a' and `b' for the bright and the faint (diffuse) region respectively. The optical spectrums were extracted over the apertures covering these two knots. The sizes of the apertures are 7.6\arcsec~and 5.9\arcsec~for the bright knot and the faint knot respectively. The selected apertures over the H$\alpha$ profile for each region are shown in Fig.~\ref{fig:02}. The redshift to the galaxy was estimated using the prominent lines of H${\alpha}$ $\lambda$6563, H${\beta}$ $\lambda$4861, H${\gamma}$ $\lambda$4340 and [O{\small{III}}] $\lambda$4959, 5007. The redshift is estimated as 0.0045$\pm$0.0002, which is in good agreement with the value provided in \citet{1999A&AS..139....1T}. The spectrums at the rest-frame wavelength are shown in Fig.~\ref{fig:03}. The calibrated spectra was de-reddened for the Galactic and internal extinction using the reddening law of \citet{1989ApJ...345..245C} with a total-to-selective extinction ratio (R$_{V}$) of 3.1. The spectrums for the bright and faint regions were first corrected for the Galactic extinction using the reddening value of E(B--V) = 0.01 in the direction of Mrk~22 estimated from \citet{2011ApJ...737..103S} recalibration of the \citet{1998ApJ...500..525S} infrared-based dust map, as implemented in NASA/IPAC Extragalactic Database (NED). The Galactic extinction corrected spectrums were then corrected for the internal extinction using the flux ratio of H${\alpha}$ and H${\beta}$ lines assuming the Case-B recombination \citep{1989SvA....33..694O,2002A&A...389..845K} with an electron temperature of $\sim$10$^{4}$ K and electron density of 100 cm$^{-3}$. For the bright nuclei, the flux ratio of f$_{H\alpha}$/f$_{H\beta}$ (Balmer decrement) is estimated as 3.80$\pm$0.49, which corresponds to E(B--V) = 0.29$\pm$0.04. For the faint region, the flux ratio f$_{H\alpha}$/f$_{H\beta}$ was found less than the theoretical expected value of 2.86. A low value of the flux ratio f$_{H\alpha}$/f$_{H\beta}$ may result from Balmer absorption \citep{1969MNRAS.145...91S} in combination with low signal-to-noise ratio of the spectrum and also sometimes due to errors in the
line flux measurements \citep{2006MNRAS.372..961K}. A low value can also result from the existence of appreciable gradients in the physical conditions such as high electron temperature or low density
in the emission region for which the theoretical ratio f$_{H\alpha}$/f$_{H\beta}$ may be less than 2.86 \citep{1980IzKry..62...54G,2009A&A...508..615L}. Previously, very low values for the f$_{H\alpha}$/f$_{H\beta}$ have been detected in several galaxies \citep{2009A&A...508..615L,2009MNRAS.396...97R}, more notably down to $\textless$ 1 in galaxies in the GAMA (Galaxy And Mass Assembly) sample derived from the SDSS data \citep{2013MNRAS.433.2764G}. Such very low values of f$_{H\alpha}$/f$_{H\beta}$ as in case of Mrk 22 are often associated with intrinsically low reddening and hence we assumed E(B--V) as zero for the faint region in Mrk 22. The prominent emission lines were identified and marked in the spectrum. These lines include the Balmer lines of Hydrogen H${\delta}$ $\lambda$4101, H${\gamma}$ $\lambda$4340, H${\beta}$ $\lambda$4861, H${\alpha}$ $\lambda$6563, He{\small{II}}~$\lambda$4686 and numerous forbidden emission lines such as [O{\small{II}}] $\lambda$3726, [O{\small{III}}] $\lambda$4363 and [O{\small{III}}] $\lambda$4959, 5007, [N{\small{II}}] $\lambda$6584, [S{\small{II}}] $\lambda$6717, 6731 and some other emission lines such as [Ne{\small{III}}] $\lambda$3868, 3967 and [Ar{\small{III}}] $\lambda$7136. The de-redenned (galactic and internal) line fluxes obtained for each line along with the equivalent widths of H${\alpha}$, H${\beta}$ and [O{\small{III}}] $\lambda$5007 lines for the two star forming regions of Mrk~22 are summarized in Table 2.


\begin{table*}\label{tab:02}
\centering
\caption{The dereddened fluxes of emission lines, equivalent widths, and derived physical parameters for the two star forming regions in Mrk~22.}
\begin{tabular}{c|c|c|c|c}
\hline
       &      & Bright region   & Faint region\\
\hline
Line  & Wavelength  & Flux & Flux\\
      &  [\AA]  & [10$^{-14}$ erg s$^{-1}$ cm$^{-2}$] & [10$^{-14}$ erg s$^{-1}$ cm$^{-2}$]\\
\hline
O{\sc II}   & 3726 &   2.63$\pm$1.18 & 0.68$\pm$0.07\\
Ne{\sc III} & 3868 &   1.15$\pm$0.21 & 0.24$\pm$0.06\\
Ne{\sc III} & 3967 &   0.86$\pm$0.13 & 0.16$\pm$0.07\\
H$\delta$  & 4101  &   0.74$\pm$0.48 & 0.15$\pm$0.09\\
H$\gamma$  & 4340  &   1.35$\pm$0.53 & 0.31$\pm$0.15\\
O{\sc III}  & 4363 &   0.26$\pm$0.04 & 0.08$\pm$0.02\\
H$\beta$   & 4861  &   2.68$\pm$0.24 & 0.72$\pm$0.09\\
O{\sc III} & 4959  &   5.50$\pm$0.56 & 1.06$\pm$0.15\\
O{\sc III} & 5007  &  16.20$\pm$1.13 & 2.41$\pm$0.27\\
He{\sc I}  & 5876  &   0.32$\pm$0.17 & 0.03$\pm$0.02\\
O{\sc I}   & 6300  &   0.06$\pm$0.02 & ---\\
N{\sc II}  & 6548  &   0.09$\pm$0.02 & 0.03$\pm$0.02\\
H$\alpha$  & 6563  &   7.64$\pm$0.46 & 0.92$\pm$0.12\\
N{\sc II}  & 6583  &   0.26$\pm$0.17 & 0.11$\pm$0.06\\
He{\sc I}  & 6678  &   0.07$\pm$0.01 & ---\\
S{\sc II}  & 6717  &   0.25$\pm$0.08 & 0.08$\pm$0.05\\
S{\sc II}  & 6731  &   0.19$\pm$0.06 & 0.06$\pm$0.04\\  
Ar{\sc III} & 7135 &   0.22$\pm$0.11 & 0.10$\pm$0.03\\
\hline
- EW (H$\alpha$) [\AA] && 1612$\pm$114 & 155$\pm$132\\
- EW (H$\beta$) [\AA] && 268$\pm$74 & 97$\pm$78\\
- EW ([O{\small{III}}]$\lambda$5007) [\AA] && 1620$\pm$452 & 294$\pm$241\\
Log ([N{\small{II}}]$\lambda$6583/H$\alpha$) && -1.5$\pm$0.3 & -0.9$\pm$0.2\\
Log ([OIII]$\lambda$5007/H$\beta$)&& 0.8$\pm$0.1 & 0.5$\pm$0.1\\
Log (R$_{23}$) && 0.96$\pm$0.05 & 0.76$\pm$0.06\\
([OIII]$\lambda$5007/[OII]$\lambda$3726) && 6.16$\pm$2.80 & 3.54$\pm$0.54\\
([NII]$\lambda$6583/[OII]$\lambda$3726) && 0.10$\pm$0.08 & 0.16$\pm$0.09\\
([OIII]$\lambda$4959+[OIII]$\lambda$5007/[OII]$\lambda$3726) &&8.25$\pm$3.73 & 5.10$\pm$0.69\\
([OII]$\lambda$3726/H$\beta$) && 0.98$\pm$0.45 & 0.94$\pm$0.15\\
([OIII]$\lambda$4959+[OIII]$\lambda$5007/H$\beta$) && 8.10$\pm$0.86 & 4.82$\pm$0.74\\
([OIII]$\lambda$4959+[OIII]$\lambda$5007/R$_{23}$) && 2.39$\pm$0.31 & 0.60$\pm$0.10\\
\hline
\end{tabular}
\end{table*}

\subsection{Radio}

The \HI 21 cm-line interferometric observations were carried out using the Giant Meterwave Radio Telescope \citep[GMRT;][]{1991CuSc...60...95S}, India. The GMRT has a hybrid configuration with both long and short baselines, which fulfill the requirements of high and low angular resolutions respectively. The observations were carried out on July 31, 2009 in a total integration time of $\sim$ 6 hours. The correlator configuration has 8 MHz bandwidth in a total of 512 spectral channels in each of the two polarizations of the L-band receiver system on the GMRT. The data reduction was performed in Astronomical Image Processing System {\small (AIPS)} with standard procedure for bad data removal and calibration. The un-resolved radio source 3C48 was used as the primary flux calibrator and the radio source 0834+555 near to Mrk~22 was used as the secondary calibrator for frequent complex gain calibrations of the GMRT antennas. The continuum subtracted, primary beam corrected, self-calibrated and de-convolved images using the {\small CLEAN} algorithm \citep{1974A&A....33..289H} are presented here. The images were made at different angular resolutions by selecting the interferometric data in different spatial frequency range. The lowest resolution ($70\arcsec \times 70\arcsec$) images were used to estimate total \HI flux and to detect extended low column density region. The highest resolution ($10\arcsec \times 10\arcsec$) images were used to obtain detailed morphologies of high column density regions in the galaxy. The \HI global profile was generated by summing flux at every velocity channel within a blotched region covering the full \HI extent of the galaxy. The zero and first order moment maps representing total \HI image and velocity field, respectively, were generated using standard procedures.  

\begin{figure*}
\centering
\includegraphics[width=18.0cm, trim=5 35 5 25, clip=true]{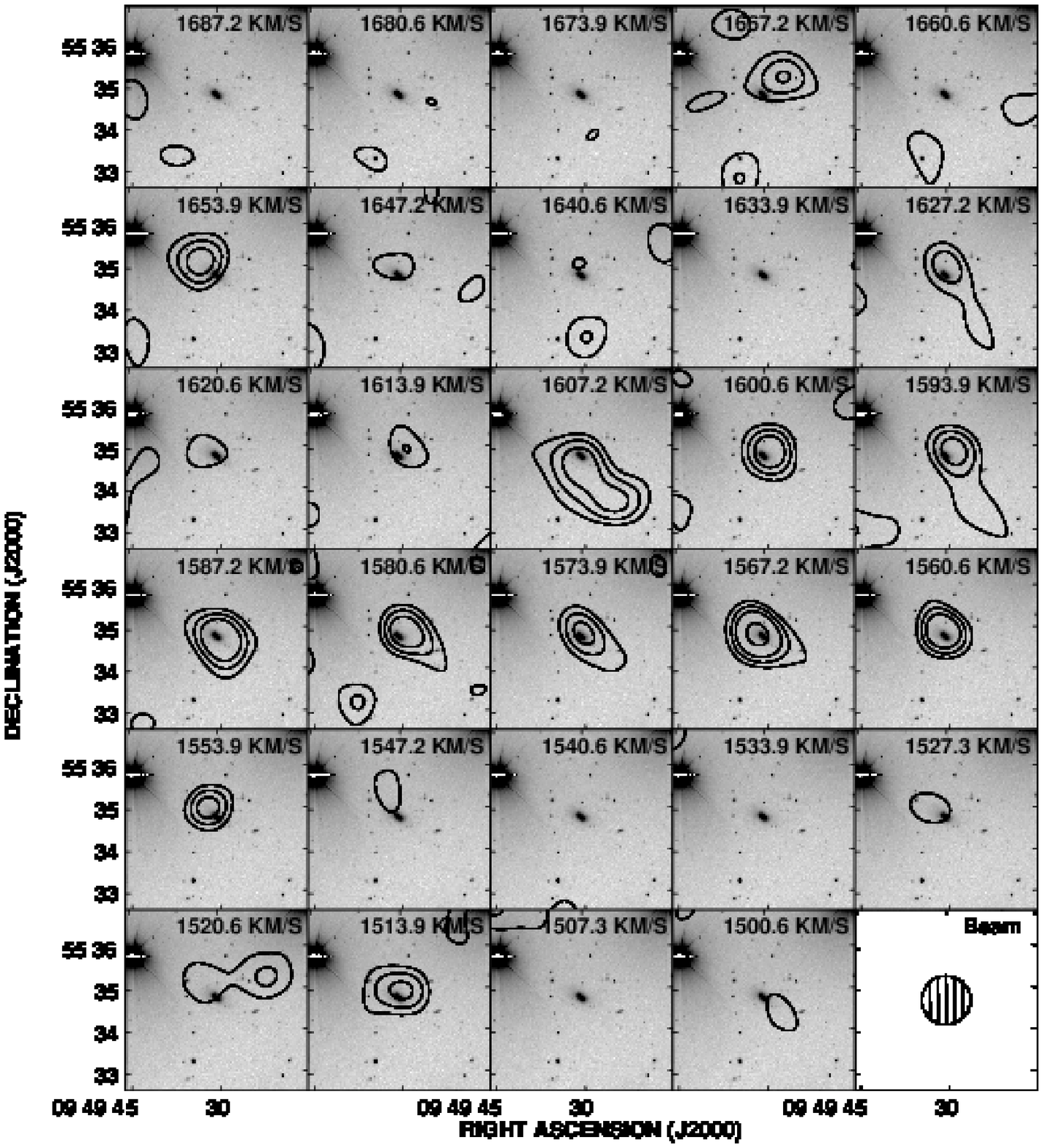}
\caption{The \HI velocity-channel contour images of Mrk 22 overlaid upon the optical image. The angular resolution is $70\arcsec \times 70\arcsec$ and the velocity resolution is 6.7 km s$^{-1}$. The \HI column density contours are shown at (1, 1.5, 2, 3, 4, ...)$\times$ 0.5 $\times$ 10$^{19}$ atoms cm$^{-2}$.}
\label{fig:04}
\end{figure*}

\begin{figure*}
\centering
\includegraphics[width=18.0cm, trim=5 35 5 25, clip=true]{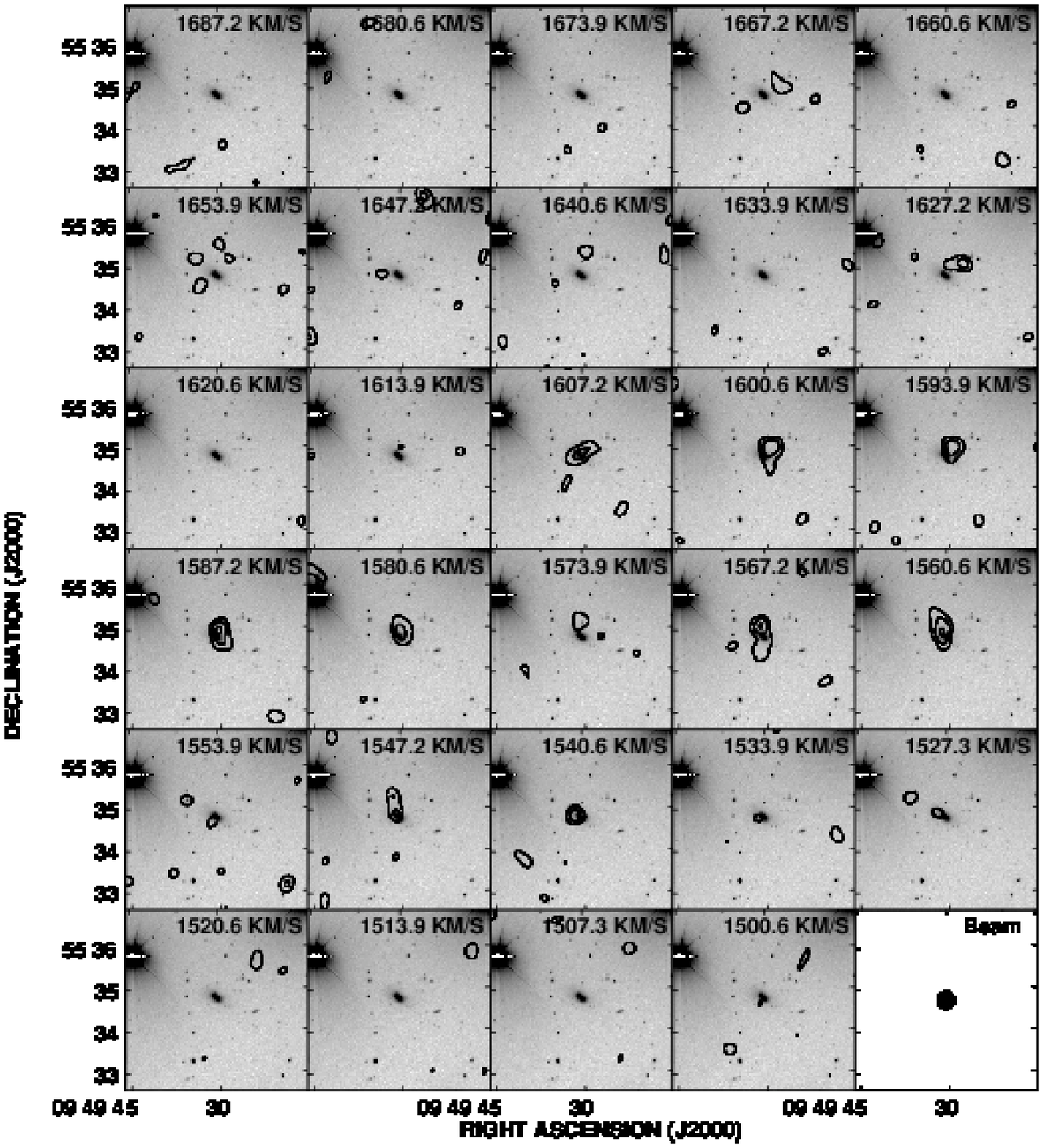}
\caption{The \HI velocity-channel contour images of Mrk~22 overlaid upon the optical image. The angular resolution is 25\arcsec $\times$ 25\arcsec and the velocity resolution is 6.7 km s$^{-1}$. The \HI column density contours are shown at (1, 1.5, 2, 3, 4, ...)$\times$ 3.7 $\times$ 10$^{19}$ atoms cm$^{-2}$.}
\label{fig:05}
\end{figure*}

The \HI velocity-channel images of Mrk~22 at resolutions of 70\arcsec $\times$ 70\arcsec and 25\arcsec $\times$ 25\arcsec are shown in Fig.~\ref{fig:04} and Fig.~\ref{fig:05} respectively. The velocity resolution is 6.7 km s$^{-1}$. The rms per channel is $\sim$ 1.9 mJy beam$^{-1}$, $\sim$ 1.2 mJy beam$^{-1}$, $\sim$ 1 mJy beam$^{-1}$ for channel images made at 70\arcsec $\times$ 70\arcsec, 25\arcsec $\times$ 25\arcsec, and 10\arcsec $\times$ 10\arcsec respectively. The \HI flux density ($S_{\HI}$) at a resolution $\theta_{a} \times \theta_{b}$ was converted to \HI column density ($N_{\HI}$) using the following relation \citep{1978ppim.book.....S}: 

\begin{equation}
N_{\HI} = \frac{1.1\times10^{21}\mathrm {cm^{-2}}}{\theta_{a}\arcsec \times \theta_{b}\arcsec} \frac{\delta v}{\mathrm{km~s^{-1}}}\frac{S_{\HI}}{\mathrm{mJy~beam^{-1}}}
\end{equation} 

\begin{figure*}
\centering
\includegraphics[width=9.0cm]{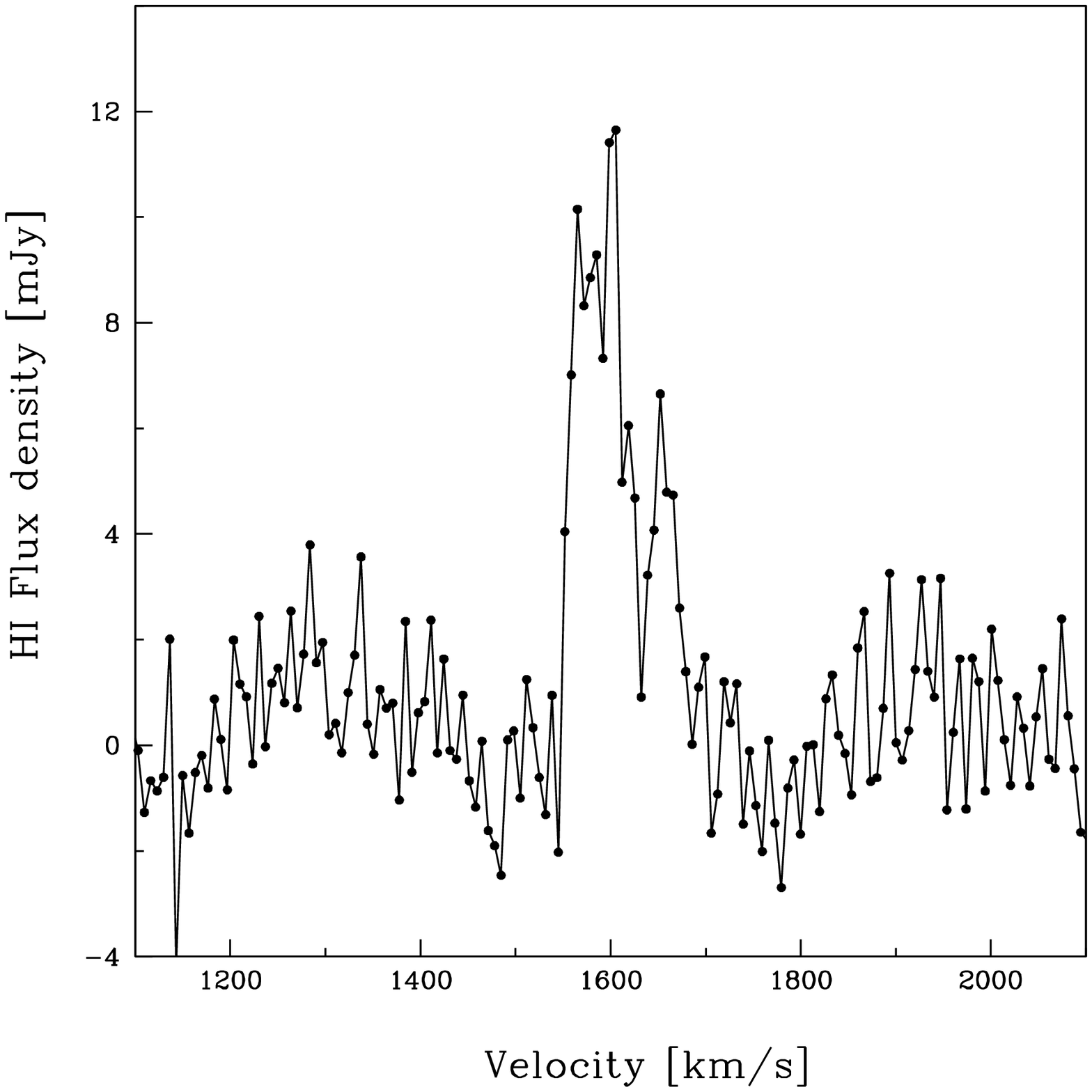}
\caption{The \HI spectrum of Mrk 22 estimated from the GMRT \HI images made at 70\arcsec $\times$ 70\arcsec resolution.}
\label{fig:06}
\end{figure*}

\begin{figure*}
\centering
\includegraphics[width=5.5cm,trim= 55 105 150 150,clip=true]{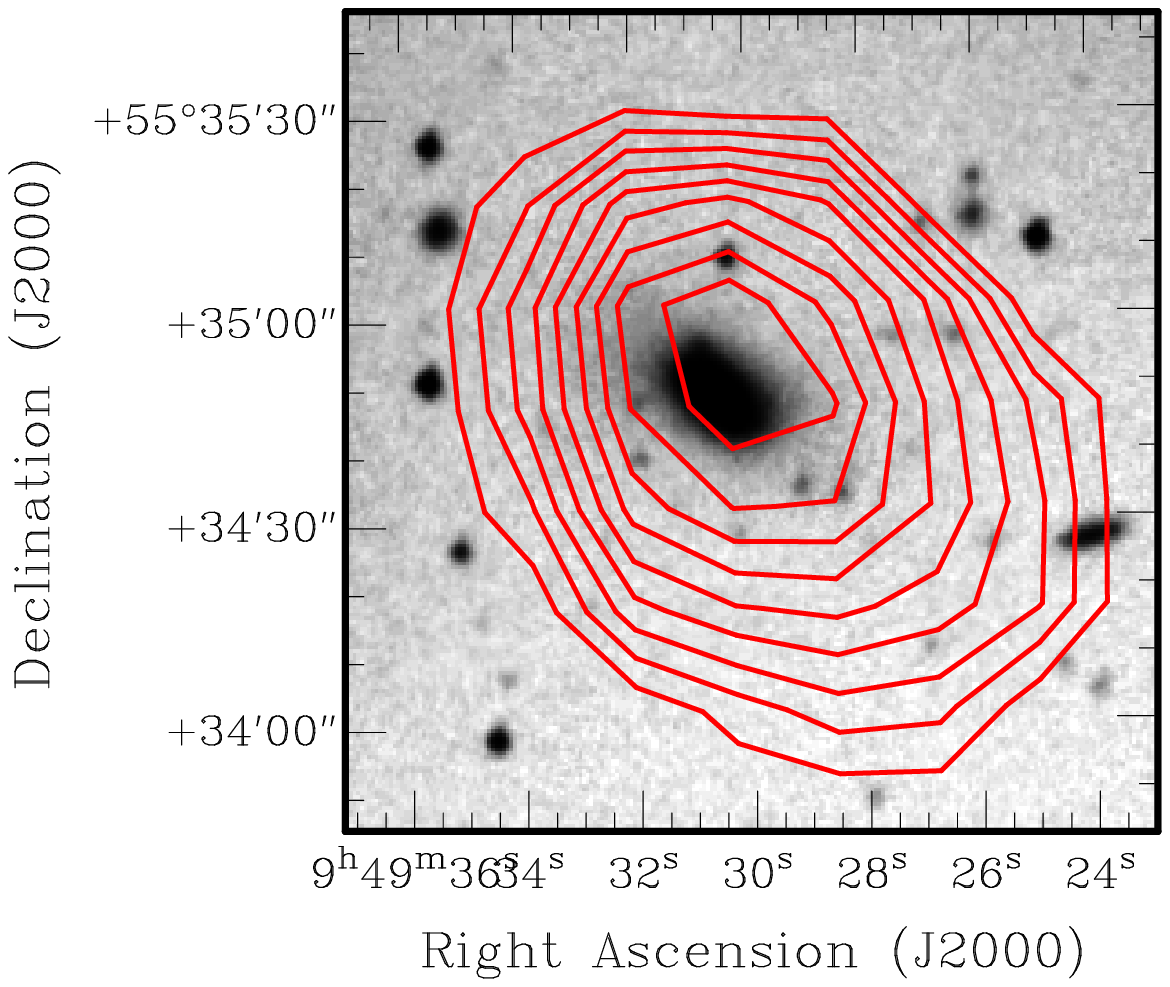}
\includegraphics[width=5.5cm,trim= 55 105 150 150,clip=true]{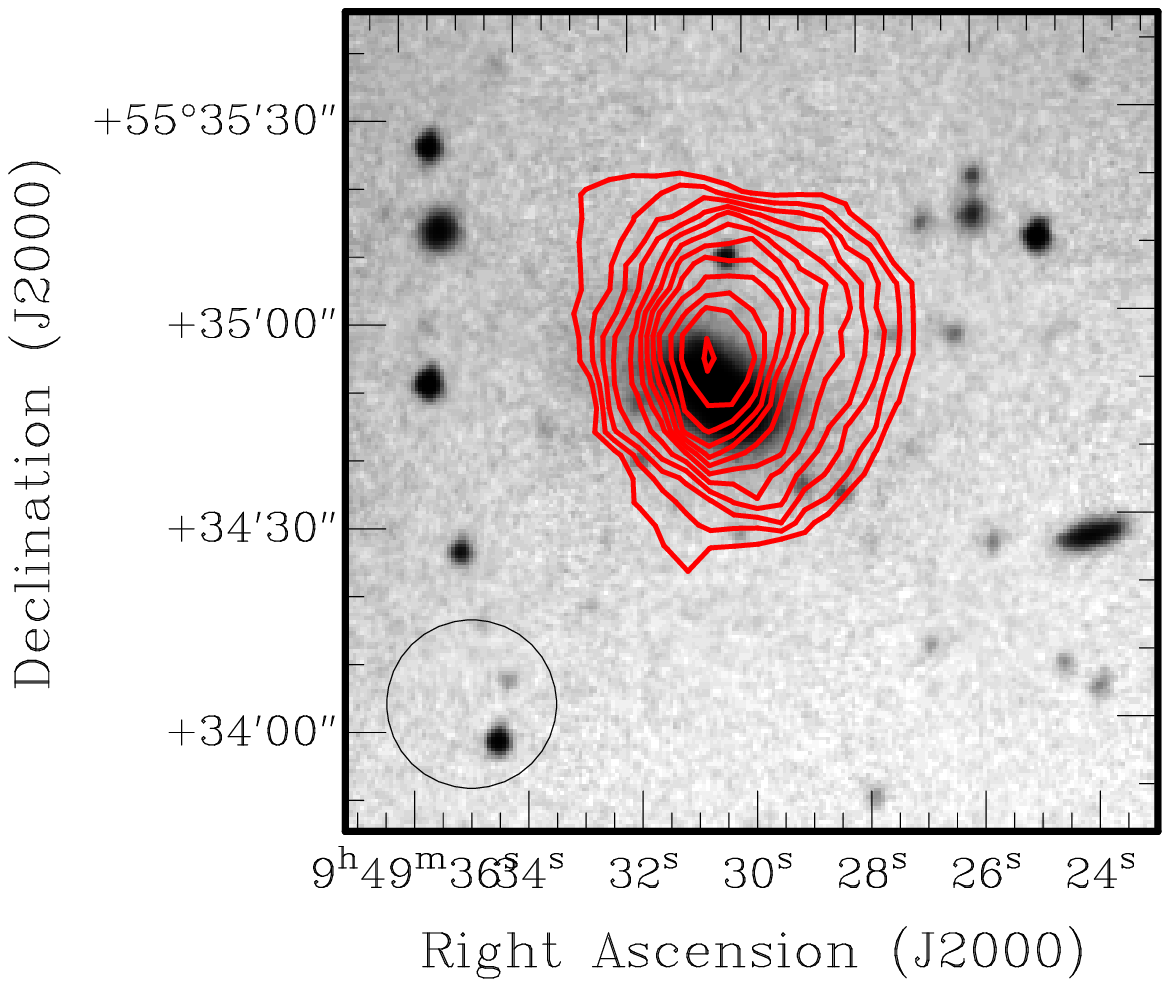}
\includegraphics[width=5.5cm,trim= 55 105 150 150,clip=true]{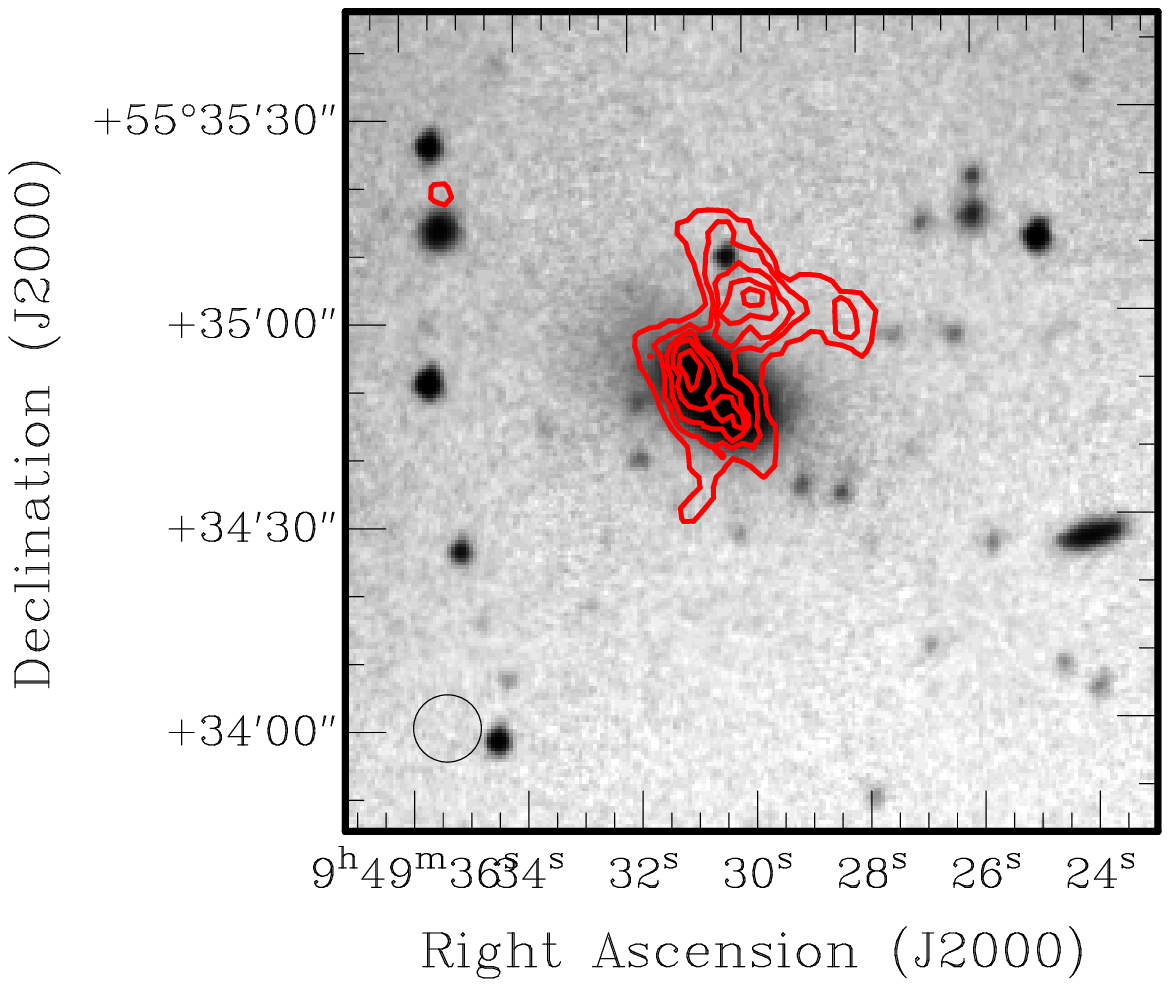}
\caption{The \HI column density contour images, overlaid upon the grayscale optical 1.3-m DFOT r-band image. The beam size in the radio images are $70 \arcsec\times 70\arcsec$ (left), $25 \arcsec \times 25\arcsec$ (middle), and $10\arcsec \times 10\arcsec$ (right). The circular beam is plotted in the middle and right panels. The \HI column density contour start at $1.1\times10^{19}$ cm$^{-2}$,  $5.3\times10^{19}$ cm$^{-2}$, $33\times10^{19}$ cm$^{-2}$ in the left, middle, and right panels respectively. The contour levels are shown as 1, 2, 3, 4... times the first level.}
\label{fig:07}
\end{figure*}

\begin{figure}
\centering
\includegraphics[width=7.5cm]{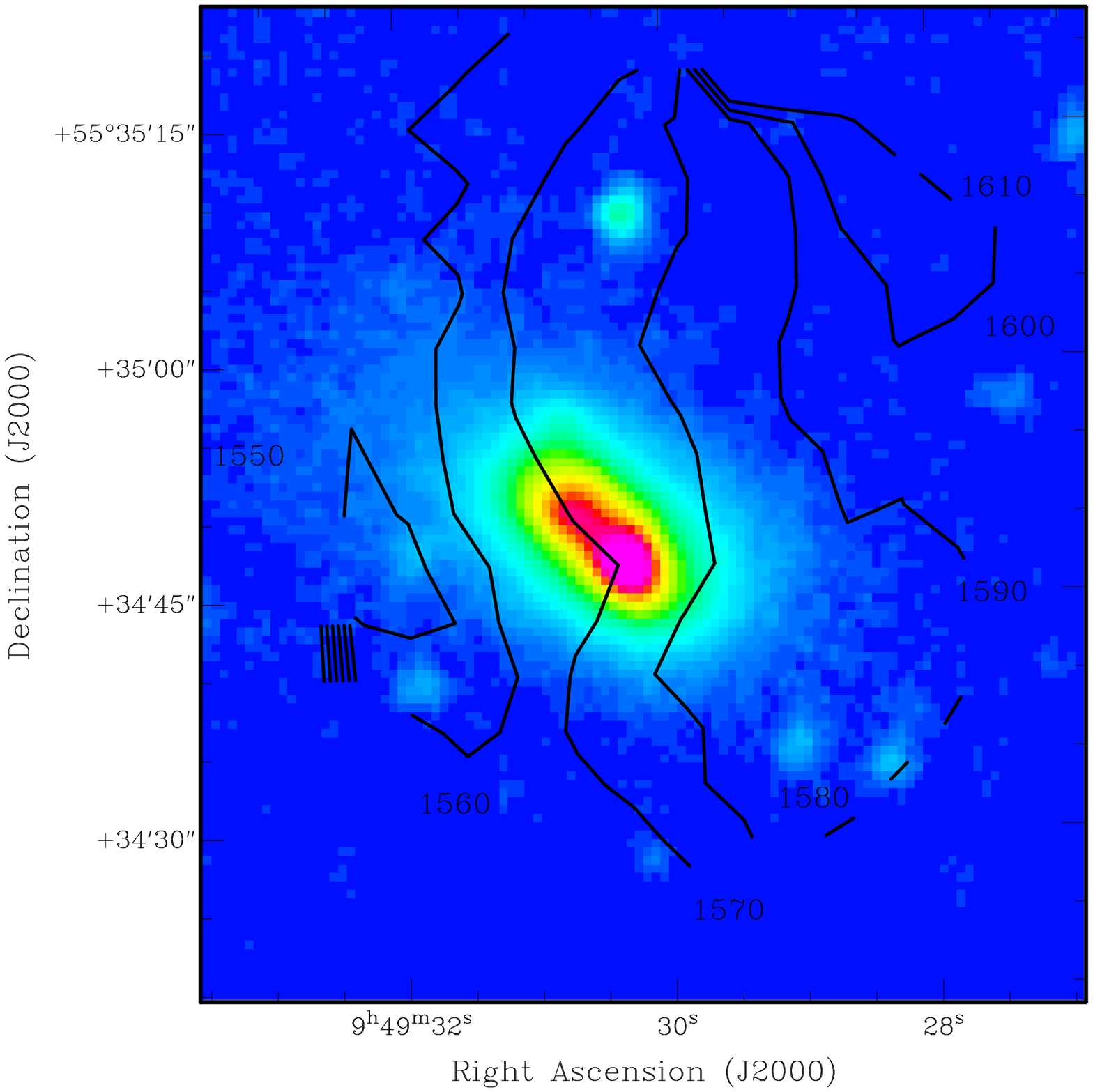}
\caption{The \HI kinematics inferred from the moment-1 image, overlaid upon the  SDSS r-band image.}
\label{fig:08}
\end{figure}

The velocities are Heliocentric and in optical definition. The velocities were corrected for the Doppler shifts caused by the projected orbital and spin motions of the Earth in the direction of the source. The \HI global profile estimated from the low resolution  channel images is shown in Fig.~\ref{fig:06}. The total \HI images at  different angular resolutions are shown in Fig.~\ref{fig:07}. The \HI velocity field in shown in Fig.~\ref{fig:08}.

\section{Analysis} 

\subsection{WR broad emission features}

\begin{figure*}
\centering
\includegraphics[width=6.5cm,height=8.5cm,angle=270]{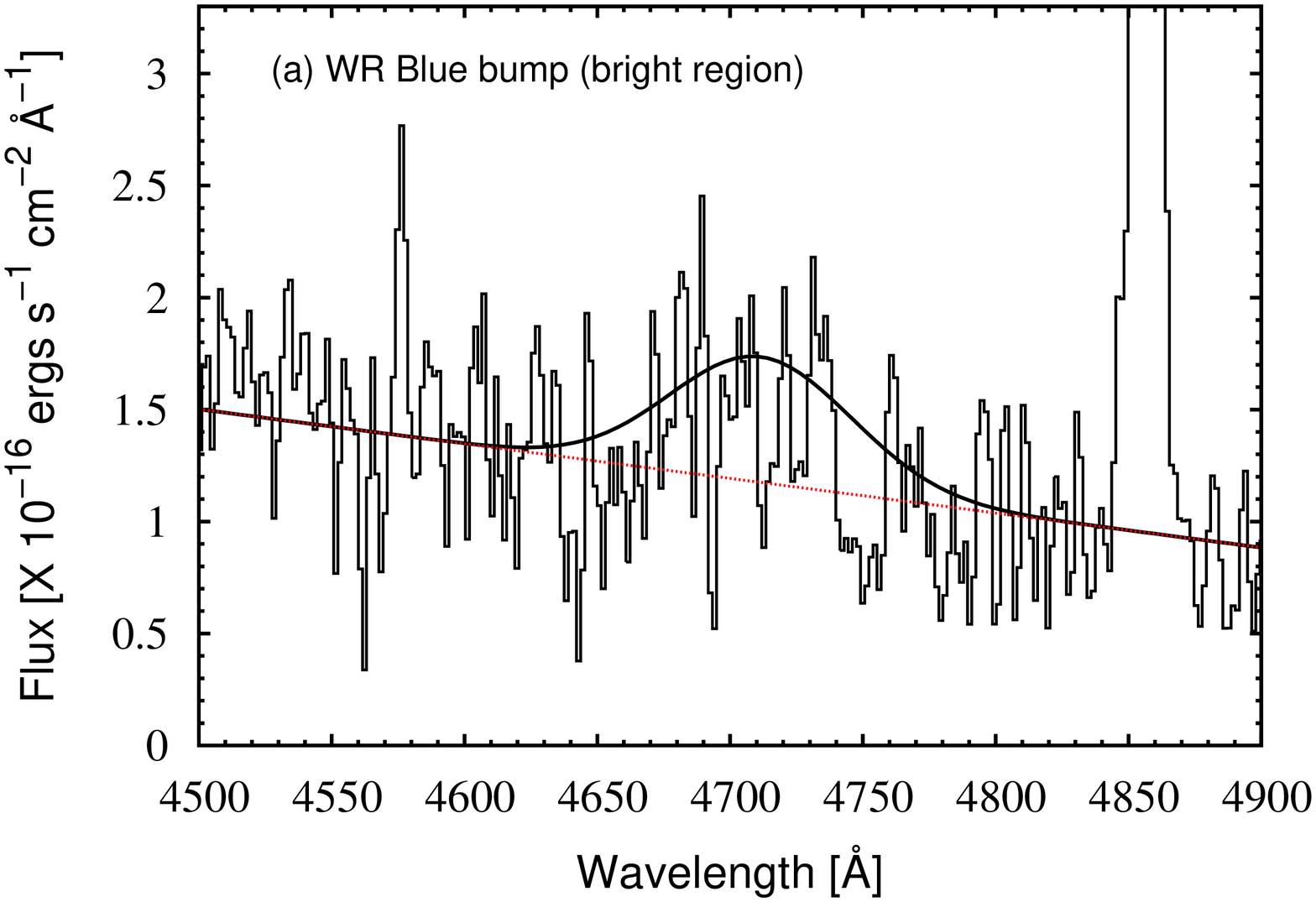}
\includegraphics[width=6.5cm,height=8.5cm,angle=270]{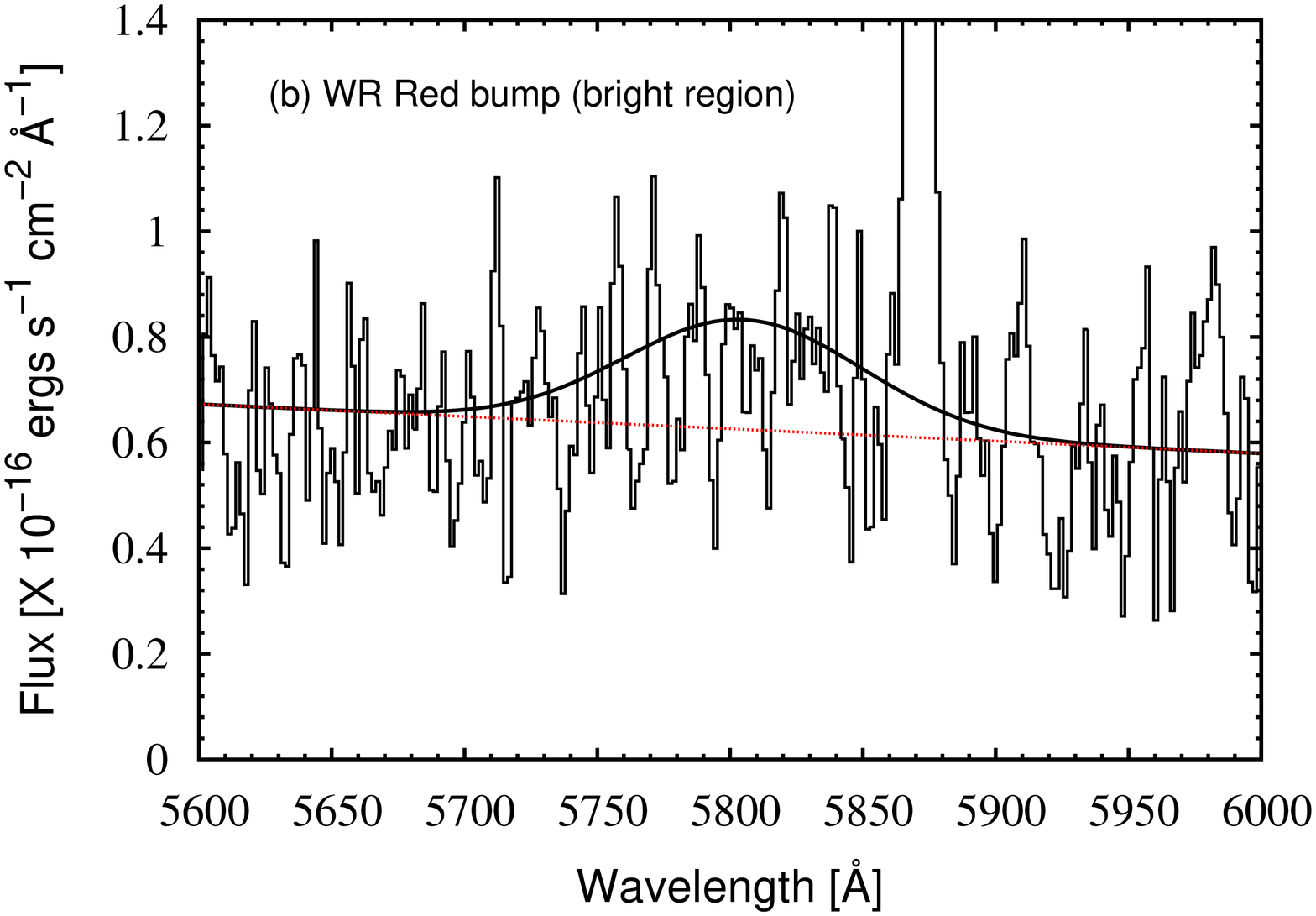}
\includegraphics[width=6.5cm,height=8.5cm,angle=270]{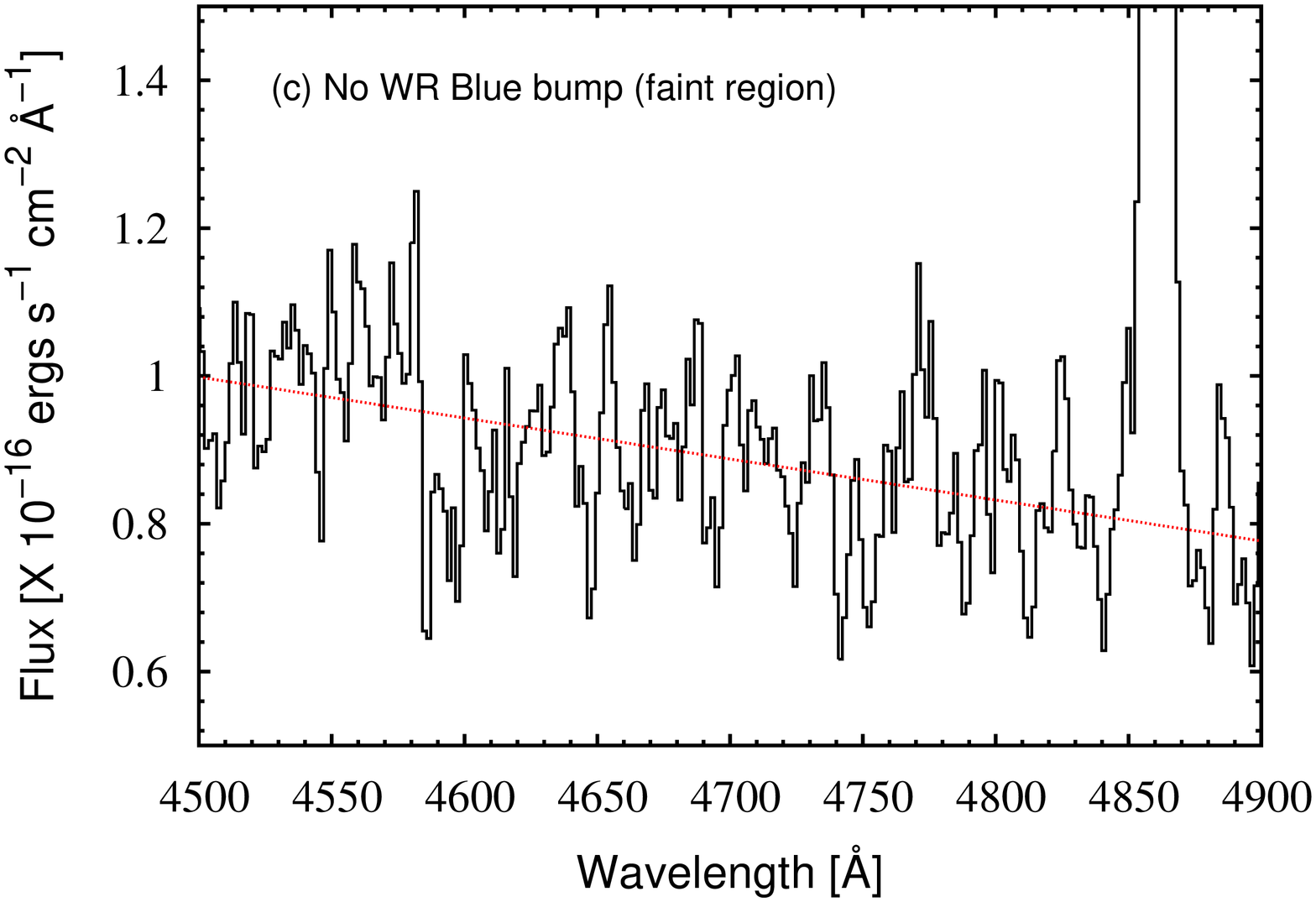}
\includegraphics[width=6.5cm,height=8.5cm,angle=270]{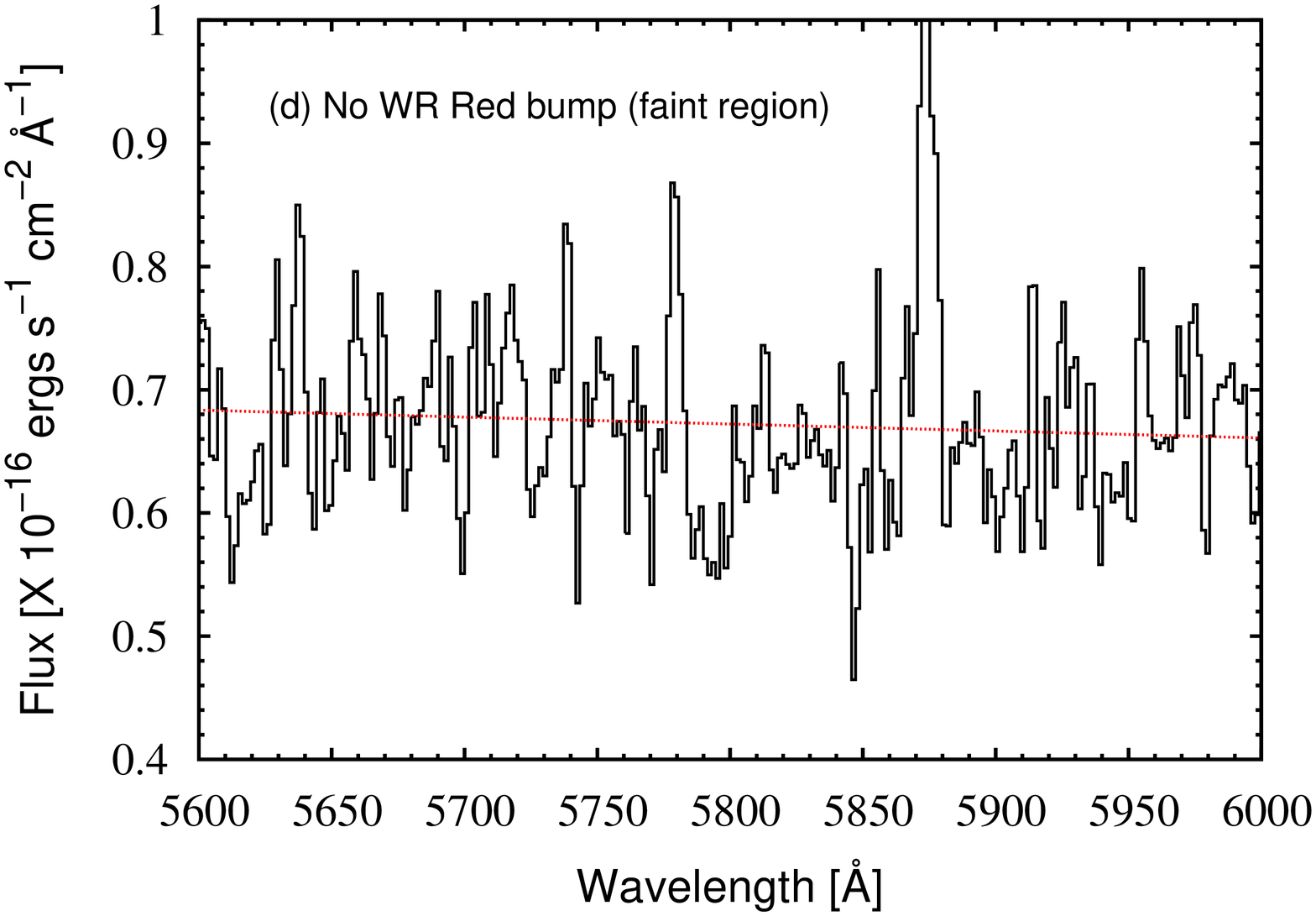}
\caption{The optical spectrum of Mrk 22 showing broad emission lines of (a) the blue WR bump and (b) the red WR bump for the bright star forming region. Similarly, panels (c) and (d) show the spectrum for the faint star forming region around the blue and red WR bump wavelengths respectively. The dotted line represents a stellar continuum fit to the spectrum, and the solid line represents a Gaussian fit to the observed spectrum.}
\label{fig:09}
\end{figure*}

The blue WR bump around 4686 \AA, which is a blend of C{\small{III}}/C{\small{IV}}~$\lambda$4650, 4658, N{\small{III}}~$\lambda$4634, 4640, Ar{\small{IV}}~$\lambda$4711, 4740 and He{\small{II}}~$\lambda$4686 emission lines, is clearly identified in the bright star forming region of Mrk 22 as shown in Fig.~\ref{fig:09}. The detection of the blue bump indicates a significant population of WR sub-types such as late-type WN (WNL) and early-type WC (WCE) stars as these lines usually appear in the spectrum due to the presence of WNL and WCE stars \citep{1998ApJ...497..618S}. Some contribution from the early-type WN (WNE) stars may also be present \citep{1998ApJ...497..618S}. However, as the lifetime of the WR stars in the WNE phase is very short in comparison to that in the WNL phase at all metallicities and masses of progenitor stars \citep{1994A&A...287..803M}, it can be safely assumed that the blending of lines over the blue WR bump is mostly due to the presence of the WNL stars. The red WR bump around C{\small{IV}}~$\lambda$5808 is also identified in the spectrum of the bright region as shown in Fig.~\ref{fig:09}. This bump appears mainly due to the presence of the WCE stars. The detected spectral features in the blue and red bump regions are consistent with the observations of \citet{2000ApJ...531..776G}. No such bumps are detected from the faint star forming region. In general, the WR phase in the most massive young stars appears after 2 to 5 Myr from their birth before the end of this phase in supernovae explosions in a very short time \citep[$t_{WR}$ $\textless$ 0.5 Myr;][]{2005A&A...429..581M}. The WR features detected for the bright star forming region in Mrk~22 are in good agreement with the starburst age ($\sim$ 4 Myr) determined in the present work (see Sect.~\ref{sec:3.5}). 

In order to quantify flux for the blue and red WR bumps, a broad Gaussian profile was fitted. The fluxes for the blue and red WR bumps are estimated as 48$\pm$16 $\times$ 10$^{-16}$ and 24$\pm$8 $\times$ 10$^{-16}$ erg cm$^{2}$ s$^{-1}$, respectively. The population of the WNL stars can be computed using the total blue bump luminosity and the luminosity of a single WNL star taken as 2.0 $\times$ 10$^{36}$ erg s$^{-1}$. The estimated number of the WNL stars is 157$\pm$59. Similarly, the number of the WCE stars is estimated as 51$\pm$19 using the red bump luminosity and the luminosity of a single WCE star taken as 3.0 $\times$ 10$^{36}$ erg s$^{-1}$. It can be seen that the WNL stars dominate in Mrk~22. The number of O-type stars can be computed assuming a Case-B recombination theory \citep{1987MNRAS.224..801H}, and using the number of ionizing photons $Q^{\rm Obs}_0$ related to the observed luminosity of the H${\beta}$ emission line using the expression: 

\begin{equation}
N({\rm O7V}) = \frac {L_{({\rm H}\beta)}}{Q^{\rm O7V}_0} = \frac {4.76\times10^{-13}Q^{\rm Obs}_0}{Q^{\rm O7V}_0}.
\end{equation}

In this analysis, an O{\small{7V}} star is taken as the representative star emitting the Lyman continuum photons, corrected for other O sub-types, at the rate of  $Q^{\rm O7V}_0$ = 1 $\times$ 10$^{49}$ s$^{-1}$ \citep{1990ApJS...73....1L,2006MNRAS.368.1822H}. The equivalent number of O{\small{7V}} stars can be converted into total number of O stars, $N({\rm O})$, assuming a time-dependent parameter $\eta_{o}$(t) defined as the ratio of the number of O{\small{7V}} stars to the number of all O{\small{V}} stars, and contribution from the WR stars using the relation:

\begin{equation}
N({\rm O})=\frac{Q^{\rm Obs}_0 - N_{\rm WR}Q^{\rm WR}_0}{\eta_0(t)Q^{\rm O7V}_0}.
\end{equation}

We assumed here $\eta_{o}$(t) $\sim$ 0.25, which is a reasonable choice for a starburst in the galaxy \citep{1998ApJ...497..618S}. The average Lyman continuum photon rate per massive star is taken as $Q^{\rm WR}_0$ = $Q^{\rm O7V}_0$ = 1 $\times$ 10$^{49}$ s$^{-1}$ given by \citet{1999A&A...341..399S}. Therefore, the total number of O-type stars in Mrk~22 is estimated as 640$\pm$318.

\subsection{Physical conditions in the ionized gas}\label{sec:3.2}

The electron temperature (T$_{e}$) and the density (n$_{e}$) of the ionized gas for two star forming regions of Mrk~22 are estimated here. The faint auroral [{\small{OIII}}] $\lambda$4363 emission line was detected along with the [O{\small{III}}] $\lambda$ 4959, 5007 emission lines in the optical spectrum. These two detections enabled us to compute the electron temperature as their relative rate of excitation strongly depends on the electron temperature. A two-zone approximation was used to estimate T$_{e}$ for the nebular regions in Mrk~22. It was assumed that T$_{e}$[O{\small{III}}] and T$_{e}$[O{\small{II}}] are the representative temperatures for the high and the low ionization potential ions, respectively. We inferred T$_{e}$[O{\small{III}}] from the diagnostic line ratio of [O{\small{III}}] I($\lambda$4959 + $\lambda$5007)/I($\lambda$4363) by using the five-level program within {\small IRAF NEBULAR} task for the emission line nebulae \citep{1995PASP..107..896S}. Once T$_{e}$[O{\small{III}}] was estimated, T$_{e}$[O{\small{II}}] was inferred using the linear relation between T$_{e}$[O{\small{III}}] and T$_{e}$[O{\small{II}}] \citep{1992AJ....103.1330G}. The values of T$_{e}$[O{\small{III}}] and T$_{e}$[O{\small{II}}] are given in Table 3 for the two star forming regions in Mrk~22. These derived temperatures are in good agreement with those measured in other nearby star forming dwarf galaxies \citep[e.g.,][]{1986MNRAS.223..811C,1994ApJ...420..576M,2004ApJ...616..752L,2008MNRAS.385..543H}. In order to measure the electron density of the ionized gas, the diagnostic line ratio of the doublet [S{\small{II}}] I($\lambda$6717)/I($\lambda$6731) was used. Since these two lines of the same ion are emitted from different levels with nearly same excitation energy, the electron density can be estimated using this line ratio. The computed electron densities are listed in Table 3.

\begin{table}\label{tab:03}
\centering
\caption{The physical conditions and chemical abundances in two prominent star forming regions in Mrk~22.}
\begin{tabular}{c|c|c}
\hline 
 Parameters & Bright region & Faint region\\
\hline 

T$_{e}$ [O{\sc III}] (K)  & 13942$\pm$811  & 19260$\pm$2715\\
T$_{e}$ [O{\sc II}] (K)   & 12759$\pm$567  & 16482$\pm$1900\\
n$_{e}$ (cm$^{-3}$)       & 94$\pm$43      & 56$\pm$29\\\\ 
12+log(O$^{++}$/H$^{+}$)  & 7.91$\pm$0.07  & 7.35$\pm$0.11\\
12+log(O$^{+}$/H$^{+}$)   & 7.19$\pm$0.21  & 6.81$\pm$0.14\\
12+log(O/H)               & 7.98$\pm$0.07  & 7.46$\pm$0.09\\\\
12+log(N$^{+}$/H$^{+}$)   & 6.02$\pm$0.22  & 5.96$\pm$0.21\\
12+log(N/H)               & 6.79$\pm$0.29  & 6.60$\pm$0.26\\
icf(N)                    & 5.91$\pm$2.67  & 4.30$\pm$1.45\\
log(N/O)                  & -1.19$\pm$0.30 & -0.86$\pm$0.27\\\\
12+log(Ne$^{++}$/H$^{+}$) & 7.40$\pm$0.09  & 6.89$\pm$0.16\\
12+log(Ne/H)              & 7.43$\pm$0.10  & 6.93$\pm$0.17\\
icf(Ne)                   & 1.07$\pm$0.07  & 1.10$\pm$0.11\\
log(Ne/O)                 & -0.56$\pm$0.12 & -0.53$\pm$0.19\\\\
12+log(S$^{+}$/H$^{+}$)   & 5.33$\pm$0.11  & 5.20$\pm$0.21\\
12+log(S/H)               & 6.13$\pm$0.24  & 5.85$\pm$0.27\\
icf(S)                    & 6.26$\pm$3.13  & 4.42$\pm$1.70\\
log(S/O)                  & -1.86$\pm$0.25 & -1.61$\pm$0.29\\\\
12+log(Ar$^{++}$/H$^{+}$) & 5.57$\pm$0.22  & 5.57$\pm$0.16\\
12+log(Ar/H)              & 5.62$\pm$0.23  & 5.57$\pm$0.16\\
icf(Ar)                   & 1.11$\pm$0.16  & 0.99$\pm$0.09\\
log(Ar/O)                 & -2.36$\pm$0.24 & -1.89$\pm$0.19\\
\hline
\end{tabular}  
\end{table}

\subsection{Chemical abundances}

We used the expressions from \citet{2006A&A...448..955I} to determine various ionic abundances. We assumed here a two-zone scheme for determining ionic abundances. T$_{e}$[O{\small{III}}] is the representative temperature for the high ionization potential ions such as O$^{++}$, Ne$^{++}$, S$^{++}$ and Ar$^{++}$ while T$_{e}$[O{\small{II}}] is the representative temperature for the low ionization potential ions such as O$^{+}$ and N$^{+}$. The ionization correction factors (ICF) of \citet{2006A&A...448..955I} were used to compute total abundances for O, N, Ne and Ar. In case of S, we followed the corrections given in \citet{1969BOTT....5....3P}. Subsequently, the log value of N/O, S/O, Ne/O, and Ar/O ratios were also computed. The resulted chemical compositions for the two star forming regions in Mrk~22 are presented in Table 3. The gas phase metallicities [12 + log(O/H)] for the bright and faint regions are estimated as 7.98$\pm$0.07 and 7.46$\pm$0.09 respectively. 

\subsection{Mass of the ionized gas}

The mass of the ionized gas was derived from the H${\alpha}$ luminosity assuming homogeneous physical conditions in the star forming regions. Using standard case-B recombination theory \citep{1974agn..book.....O}, the mass of the ionized gas can be written as \citep{1994A&AS..105..341G,1996A&AS..120..463M}:

\begin{equation} 
M_{HII} = 2.33 \times 10^{3} (\frac{L_{H\alpha}}{10^{39}})(\frac{10^{3}}{n_{e}}) M_{\odot}
\end{equation}

\noindent where L$_{H\alpha}$ and n$_{e}$ are the H${\alpha}$ luminosity and the electron density, respectively. The mass of the ionized gas for the bright and the faint region are estimated as 1.2 $\pm$ 0.5 $\times$ 10$^{5}$ and 2.5 $\pm$ 1.4 $\times$ 10$^{4}$ M$_{\odot}$ respectively.
 
\subsection{Age of the starburst}\label{sec:3.5}

\begin{figure*}
\centering
\includegraphics[width=6.5cm,height=8.5cm,angle=270]{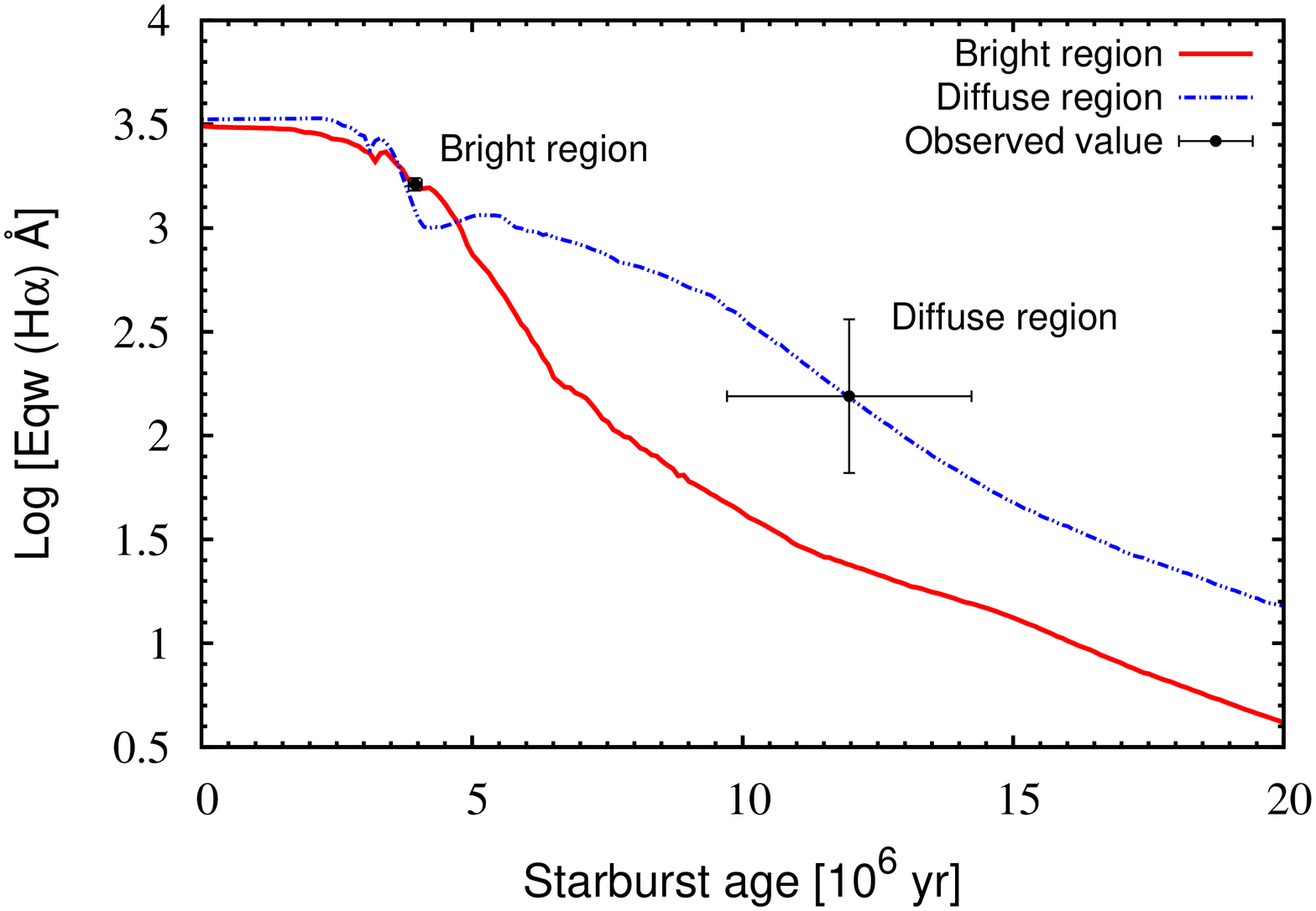}
\includegraphics[width=6.5cm,height=8.5cm,angle=270]{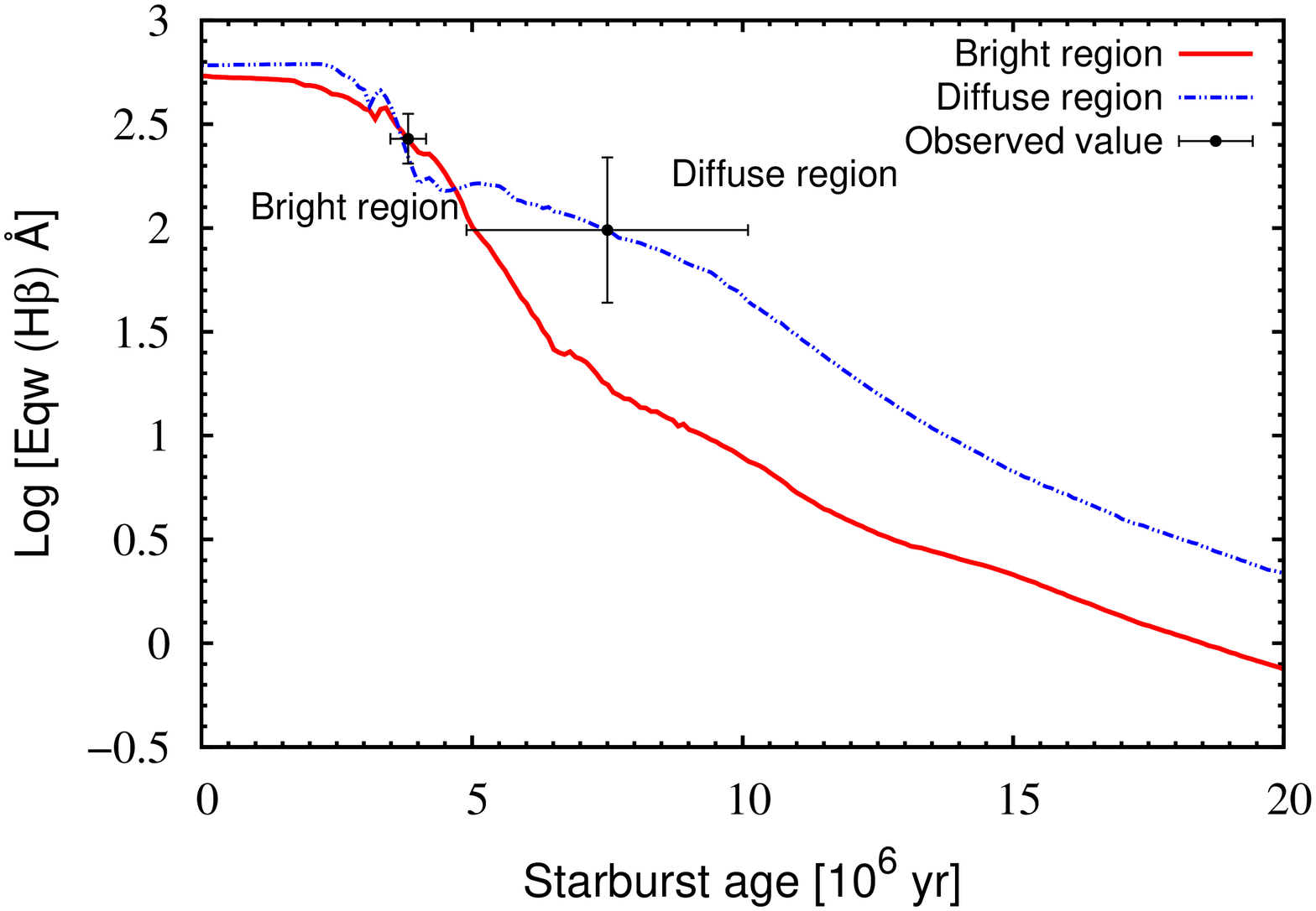}
\caption{The computed tracks for H$\alpha$ (left) and H$\beta$ (right) EW versus age of the starburst in Mrk~22. The points (solid circle) with error bar represent the observed EW and the estimated age based on these tracks.}
\label{fig:10}
\end{figure*}

An age of the most recent starburst can be predicted from H${\alpha}$ and H${\beta}$ equivalent widths (EW) as the EW decreases with time in a well defined manner \citep{1995ApJ...454L..19L,2000AJ....119.2146J}. We used the Starburst99 model provided by \citet{1999ApJS..123....3L} to predict the age of the most recent starburst in Mrk~22. The Padova stellar evolutionary model with asymptotic giant branch (AGB) evolution was fitted to obtain equivalent width track of the starburst by assuming the Salpeter initial mass function (IMF) with lower and upper mass limits as 0.1 M$_{\odot}$ and 100 M$_{\odot}$ respectively. The model used a total fixed mass of 10$^{6}$ M$_{\odot}$ with an instantaneous star formation law. The metallicities as $\sim$8 and $\sim$7.5 were used for the bright and the faint region, respectively. The EW tracks for the H$\alpha$ and the H$\beta$ lines correspond to the two star forming regions in Mrk~22 are shown in Fig.~\ref{fig:10}. For the bright region, the starburst age is estimated as 3.9$\pm$0.1 and 3.8$\pm$0.3 Myr using H${\alpha}$ and H${\beta}$ EWs respectively. For the faint region, the starburst age is estimated as 12$\pm$2 and 7.5$\pm$2.6 Myr using H${\alpha}$ and H${\beta}$ EWs respectively. We took a mean age of the starburst for the bright region as 3.9$\pm$0.3 Myr, and similarly for the faint region as 10$\pm$2 Myr. 

\subsection{\HI morphology and kinematics}

The \HI channel images shown in Figs.~\ref{fig:04} and \ref{fig:05} mark detection of \HI emission from Mrk~22 in the velocity range 1670 km s$^{-1}$ to 1510 km s$^{-1}$ with a total width of 160 km s$^{-1}$. The \HI morphology as indicated by the moment-0 image (Fig.~\ref{fig:07}) of Mrk~22 is irregular with an indication of tidal interaction. Large high column density \HI plume is seen outside the galaxy along the north direction. An extra component of the diffuse \HI emission is also seen extending towards the south-east direction between 1607 and 1593 km s$^{-1}$. The \HI kinematics as inferred from the moment-1 image (Fig.~\ref{fig:08}) is disturbed as it does not show a regular and smooth rotation as expected in late-type galaxies. The channel images show dis-continuity in the \HI emission between 1667 km s$^{-1}$ and 1633 km s$^{-1}$, and between 1527 km s$^{-1}$ and 1507 km s$^{-1}$. Since the emission is seen projected in the sky-plane, we speculate that these emission regions are tidal tails and bridges oriented roughly along the line of sight. The galaxy appears in an advanced stage of merger. The total \HI flux integral and \HI mass are estimated as 0.87$\pm$0.05 Jy km s$^{-1}$ and 0.88$\pm$0.05 $\times$ 10$^{8}$ M$_{\odot}$ from the lowest resolution $70\arcsec \times 70\arcsec$ images. The peak column density in Mrk~22 is $\sim1.7 \times 10^{21}$ cm$^{-2}$ at a resolution of $10\arcsec \times 10\arcsec$ at the locations of the star forming regions identified in the H${\alpha}$ image and in the \HI plume outside the galaxy. 

\subsection{Radio continuum} 

\begin{figure}
\centering
\includegraphics[width=8.0cm,height=8.0cm]{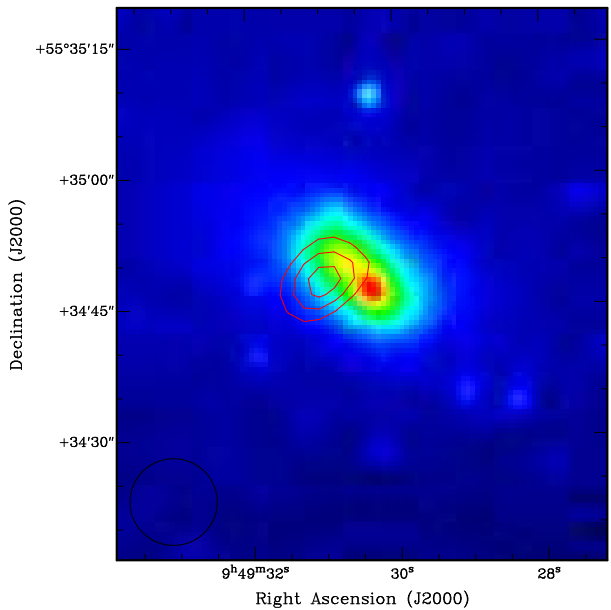}
\caption{The 1.4 GHz radio continuum contours plotted over the DFOT r-band image of Mrk~22. The black circle in the lower right corner indicates the radio beam size of $10\arcsec \times 10\arcsec$. The contour levels are at 0.5, 0.6, and 0.7 mJy beam$^{-1}$.}
\label{fig:11}
\end{figure}

The radio continuum from galaxies originates from two processes, namely, thermal free-free emission localized within the H{\small{II}} regions and diffuse non-thermal synchrotron emission from the relativistic electrons moving in the galactic magnetic field \citep{1992ARA&A..30..575C}. The cosmic electrons are accelerated to relativistic speeds in supernovae explosions of the massive stars. The non-thermal emission is the dominant component ($>80$\%) of the total radio continuum from normal spiral galaxies at 1.4 GHz. The 1.4 GHz radio continuum contour image from the Faint Images of Radio Sky at Twenty-centimeters (FIRST) survey using the Very Large Array (VLA) is shown in Fig.~\ref{fig:11}. The noise in the FIRST image is $\sim0.13$ mJy beam$^{-1}$ at its native resolution of nearly $5\arcsec \times 5\arcsec$. The image shown in Fig.~\ref{fig:11} is convolved with a larger beam size of $10\arcsec \times 10\arcsec$. The convolution to a larger beam helps in detecting diffuse emission. The noise in the convolved image is marginally higher at 0.15 mJy beam$^{-1}$. Some radio continuum emission is detected within the optical extent of the galaxy. However, the peak of the radio emission is offset from the peak of the \Ha emission. Total radio flux is estimated as $0.9\pm0.3$ mJy. It is challenging to associate this radio emission with any of the current star forming region in the galaxy. There is a possibility that this radio source is some background radio source, which is not associated with Mrk 22. It is nonetheless interesting to note the absence of radio emission from the bright star forming region. The expected 1.4 GHz radio flux can be estimated from the \Ha flux assuming that both radio (thermal and non-thermal) and \Ha trace massive star formation. The current massive SFR in galaxies can be estimated using the following relations \citep[cf.,][]{1998ApJ...498..541K,2002AJ....124..675C}: 

\begin{equation}
\frac{SFR_{H\alpha}}{\rm M_\odot~yr^{-1}} = 9.45\times10^{8} \frac{D^{2}}{\rm Mpc^{2}} \frac{f_{H\alpha}}{\rm erg~cm^{2}~s^{-1}}
\end{equation}

\begin{equation}
\frac{SFR_{1.4 GHz}}{\rm M_\odot~yr^{-1}} = 2.63\times10^{-5} \frac{D^{2}}{\rm Mpc^{2}} \frac{f_{1.4 GHz}}{\rm mJy}
\end{equation}

By equating SFRs estimated from \Ha and radio, we obtain a relation between expected radio flux and the observed \Ha flux as: 

\begin{equation} 
\frac{f_{1.4GHz - total}}{\rm mJy} \sim 3.6 \times 10^{13}~\frac{f_{H\alpha}}{\rm erg~cm^{-2}~s^{-1}}
\end{equation}

It may be noted that the above relation is only an approximate relation and actual values can differ by a factor up to 2 in galaxies. The \Ha flux estimated from the \Ha narrow-band images is 9.4$\pm$1.2 $\times$ 10$^{-14}$ and 1.3$\pm$0.2 $\times$ 10$^{-14}$ erg cm$^{2}$ s$^{-1}$ for the bright and faint regions respectively. The SFR is estimated as 0.05$\pm$0.01 M$_{\odot}$ yr$^{-1}$ and 0.007$\pm0.001$ M$_{\odot}$ yr$^{-1}$ for the bright and faint regions respectively. The expected radio (thermal and non-thermal) flux is $\sim$ 3.4 mJy and $\sim$ 0.5 mJy for the bright and faint regions respectively. It is clear that the expected radio flux is much higher than the detection sensitivity in the FIRST images. We also looked for 1.4 GHz radio emission in another survey at 1.4 GHz, namely NRAO VLA Sky Survey (NVSS) which has higher sensitivity for diffuse extended sources at an rms of 0.45 mJy beam$^{-1}$ with a beam size of $45\arcsec \times 45\arcsec$. No detection was made in the NVSS images also. It is also worthwhile to point out here that the expected thermal radio continuum typically at a level of $10 - 20$ per cent of the total radio emission could be below the detection limit of the FIRST and NVSS images. 

\subsection{Ionization mechanism}

\begin{figure}
\centering
\includegraphics[width=6.5cm,height=8.5cm,angle=270]{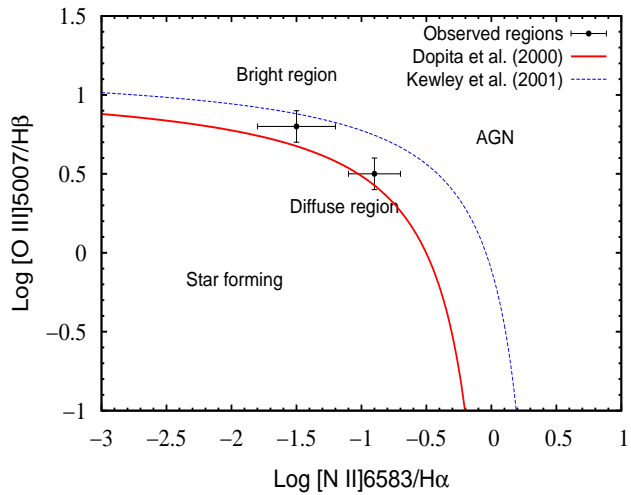}
\caption{A comparison of the observed line flux ratios obtained from the two star forming regions in Mrk~22 along with the diagnostic diagrams proposed by Dopita et al. (2000) and Kewley et al. (2001).}
\label{fig:12}
\end{figure}

The nature of the ionizing source can be inferred using the quantitative classification scheme proposed by \citet{1981PASP...93....5B} using combinations of line ratios. Subsequently, a similar diagnostic diagram was proposed by \citet{2000ApJ...542..224D} and \citet{2001ApJ...556..121K}. The BPT diagnostic scheme uses the line ratios of [O{\small{III}}] $\lambda$5007/H${\beta}$ and [N{\small{II}}] $\lambda$6583/H${\alpha}$. Various optical line ratios are listed in Table~\ref{tab:02} for the two star forming regions in Mrk~22. The locations of two star forming regions in Mrk~22 in BPT diagram are indicated in Fig.~\ref{fig:12}. The line ratios indicate that the primary and dominating source of ionization is photo-ionization in both the regions. 

\subsection{Local galaxy environment}

It is worth examining the local environment and galaxy density in the vicinity of Mrk~22 in view of tidal interaction identified in the \HI images. We searched for neighbors within a radius of 2 degree, within a velocity range of 1350 km s$^{-1}$ and 1750 km s$^{-1}$, i.e, nearly $\pm$ 200 km s$^{-1}$ from the  recession velocity of Mrk~22. The reasons for selecting this parameter space are the following. A galaxy can travel up to $\sim$ 1 Mpc over a time of 10 Gyr (age of galaxy) with a velocity of 200 km s$^{-1}$ in the IGM, which is typical velocity dispersion in groups of galaxies. At the distance to Mrk~22, a 1 Mpc region will be roughly of 2 degree extent. A total of 7 galaxies are listed in the Nasa Extragalactic Database (NED) within this parameter space, implying an average galaxy density of 2 Mpc$^{-2}$. The identified galaxies are KUG 0942+551, KUG 0951+568, SBS 0941+569B, SBS 0943+543, SBS 0940+544/KUG 0940+544, and UGC 5369, which have velocities in a narrow range of 1500 to 1650 km s$^{-1}$. It appears that Mrk~22 belongs to a small group of galaxies. Interestingly, except UGC 5369 (S0?), all other galaxies are star forming dwarf galaxies in this region as they appear blue in the SDSS images. SBS 0940+544 is also a BCD galaxy with an indication of continuous star formation during the past several giga-years \citep{2001A&A...378..756G}. 

\section{Discussion}

\subsection{Star formation and abundances}

The young starburst inferred by the detections of high ionization emission line of He{\small{II}}~$\lambda$4686 and the blue WR bump in this work and previous works \citep{2000ApJ...531..776G,2008A&A...485..657B} is confirmed by the age estimates made here for the bright and faint regions in Mrk 22 as $\sim$ 4 Myr and $\sim$ 10 Myr respectively. Unlike previous works, we carried out abundance analysis for both the regions separately. We found an appreciable metallicity difference of $\sim$ 0.5 dex between the bright region [12 + log (O/H) $\sim$ 8] and the faint region [12 + log(O/H) $\sim$ 7.5]. The separation between two regions is $\sim$ 0.6 kpc. Typical metallicity gradients in normal spiral galaxies have been found between -0.009 dex kpc$^{-1}$ and -0.231 dex kpc$^{-1}$, with an average gradient of -0.06 dex kpc$^{-1}$ \citep{1994ApJ...420...87Z}. The observed metallicity difference between the two regions in Mrk  22 is too large to be explained as a normal galactic metallicity gradient. The  chemical composition as measured from the gas-phase metallicity [12 + log(O/H)] shows various degree of spatial variations in different types of dwarf galaxies. For instance, shallow gradient in metallicity is seen in SBS 0335-052 \citep{2006A&A...454..119P} while no significant variations were seen in Mrk~35 \citep{2007ApJ...669..251C}. A study on a large sample indicates that normal BCD galaxies are chemically homogeneous \citep{1996ApJ...471..211K,2006A&A...454..119P,2008A&A...477..813K,2009ApJ...707.1676C,2009MNRAS.398..949P,
2011MNRAS.412..675P,2011MNRAS.414..272H,2012MNRAS.423..406G,2013AdAst2013E..20L}. On the other hand, the metallicity of extremely metal-poor galaxies is usually not homogeneous within the galaxy, with the low metallicity seen in regions of intense star formation \citep{2006A&A...454..119P,2009ApJ...690.1797I,2011ApJ...739...23L,2013ApJ...767...74S,2014ApJ...783...45S,
2015ApJ...810L..15S}. However, large metallicity gradients are not common in dwarf galaxies. The simplest explanation for large metallicity difference in a single system is a recent merger of two galaxies with different metallicity. In a few cases, significantly large metallicity differences between star forming regions in dwarf galaxies were seen and understood in terms of recent tidal interactions or mergers \citep{2004A&A...428..425L,2004ApJS..153..243L,2006A&A...449..997L,2009A&A...508..615L,2010A&A...517A..85L}. The evolution in terms of metallicities in interacting dwarf galaxies is fairly complex as it can depend on various factors such as mixing of metals with the ISM, possible outflows of metals, and inflow of metal-poor gas in tidally interacting systems. 

The winds from the most massive stars and the high rate of supernovae in the young star-bursting region are expected to enrich the ISM. Various studies have addressed this aspect in BCDs \citep{1979A&A....80..252C,1979A&A....80..234C,1981A&A....99...97M,1983A&A...120..113M,1992A&A...264..105M,
1993fces.conf..173M,1986ARA&A..24..329C,2002ApJ...567..532H,2009A&A...508..615L,2010A&A...517A..85L,
2013ApJ...764...44Z}. The $\alpha $ elements-to-oxygen abundance ratios in two star forming regions in Mrk~22 do not show any significant trends with observed metallicity, in agreement with results presented in \citet{2006A&A...448..955I}. Although, there is no appreciable difference in N/O ratio in two regions within the errors in the estimates, the lower value of -0.86$\pm$0.27 estimated for the fainter and the older region is indicating some N-enrichment in this slightly evolved star forming region. The N/O ratio is widely discussed in star forming dwarf galaxies. The N/O ratio remains fairly constant at low metallicities [12 + log(O/H) $\textless$ 8.0] and tend to increase at higher metallicities \citep{2006A&A...448..955I,2012A&A...546A.122I,2014A&A...569A.110L,2014MNRAS.445.2061L} with a significantly large scatter at intermediate metallicities. This observed behaviour of N/O ratio in galaxies is understood in terms of two sources of enrichment - (i) primary production of metals in winds of massive stars and supernovae in star-bursting galaxies, and (ii) secondary production from intermediate low mass old stars in higher metallicity systems. The relative elemental abundances in gas-phase depend on the age of star forming regions, metallicity, amount of dust and selective depletion of certain elements in dust grains. The N-enrichment from the winds of the massive stars can be highly significant as for an instance, a 60 M$_{\odot}$ star will release half of its mass in winds within the first 5 Myr. The selective depletion of oxygen into dust grains can cause apparent increase in N/O ratio with decreasing EW(H$\beta$) \citep{2006A&A...448..955I,2008A&A...485..657B,2010A&A...517A..85L}. As star forming regions evolve with time, the EW decreases. It is important to note that N/O enrichment can also depend on the chemical evolution during an episode of massive and sudden accretion of metal-poor gas in interactions or mergers with gas cloud or small gas-rich companions \citep[e.g.,][]{2005A&A...434..531K,2008MNRAS.385.2181F}. The enrichment can further depend on the observed spatial scale over which the population of the WR stars is distributed \citep{2011A&A...532A.141P}. Several studies have indicated that explanations for the observed N/O ratio are not straight forward as it may include several  mechanisms such as time delays on chemical enrichment, effect of gas flows, and variations in the star formation histories \citep[e.g.,][]{1990PASP..102..230G,1992A&A...260...58P,1999MNRAS.306..317K,2006ApJ...636..214V,2006MNRAS.372.1069M,
2012MNRAS.421.1624P}. Overall, the N/O and other elemental abundance ratios seen in Mrk~22 are similar to the values seen in other dwarf galaxies, and also consistent with the results of \citet{2006A&A...448..955I}. 

There is a possibility that the newly synthesized heavy elements in younger regions remain in hot phase and are not well mixed with the ISM. In such a scenario, the observed gas-phase metallicity does not necessarily come from the present burst. It is also shown by simulations that metals in hot phases are preferentially ejected away from their parent site of formation and may even be completely ejected from low mass dwarf galaxies \citep{1999ApJ...513..142M} in supernovae explosions. The tidal interactions generating inflows and outflows of gas can allow mixing of metals with the ISM. It should also be noted that in very young starbursts with WR features, supernovae explosions may not have taken place and hence metal ejection process may not have started. However, the metals produced in the winds of the massive stars should be observable in this phase. The time scales of ejection and mixing, and mass of the galaxy become important here. Our aim of discussions here is to point out that the mixing of metals with the ISM during outflows and inflows becomes difficult to understand in presence of tidal interactions and intense star burst. The observed metallicity in dwarf galaxies can therefore be a function of time stage of star-bursts and may vary over short time scales of a few 100 Myr or so. This conjecture can also be supported by large scatter observed in metallicity of star forming dwarf galaxies at a given stellar mass or blue luminosity \citep{2004ApJS..153..243L, 2004A&A...428..425L, 2008A&A...491..131L, 2009A&A...508..615L, 2010MNRAS.401..759J, 2010A&A...517A..85L}.

\subsection{Galaxy environment and tidal interactions}

The \HI morphology indicates that Mrk~22 is a tidally interacting system in an advanced stage of merger. We also noticed that the northern side of Mrk~22 has boxy isophotes in optical images. The boxy isophotes are expected in equal mass merger \citep{1999ApJ...523L.133N}. Some early-type transiting dwarf galaxies believed to be merger remnant also show similar features \citep{2007ApJ...655L..29D}. The merger in Mrk~22 is taking place most likely with another dwarf or low mass system as there is no massive galaxy found in the vicinity. Mrk~22 is residing in a group environment. With an estimated galaxy density of about 2 Mpc$^{-2}$ around Mrk~22 and an interaction rate of roughly one per 10 Gyr, it is obvious that tidal interactions leading to intense starbursts are not very common in this region. 

It is known that star-forming dwarf galaxies reside in low density environments, with a majority having no massive galaxy in the close vicinity \citep{2011ApJ...743....8W}. The star formation triggering in dwarf galaxies is believed to both internal such as stochastic self propagating star formation mechanism \citep{1980ApJ...242..517G} or external such as tidal interactions. The later hypothesis has strong observational support as many dwarf galaxies show nearby companion or ongoing interaction with another dwarf or \HI cloud \citep{2001A&A...371..806N,2001A&A...374..800O,2001ApSSS.277..445P,2008MNRAS.388L..10B,
2008MNRAS.391..881E,2010MNRAS.403..295E,2016MNRAS.462...92J}. Several simulations have shown that major interactions can drive starbursts and gas inflow from the outskirts of the \HI disks \citep[e.g.,][]{2010ApJ...710L.156R}. Studies suggest that star formation in dwarf galaxies is likely to be quasi-continuous with episodic bursts of star formation separated by relatively long phases of quiescence \citep[e.g.,][]{1991ApJ...370...25T,1995A&A...303...41K,1999A&A...349..765M,2000ApJ...539..641T,2001AJ....121.2003V}. The Hubble space telescope observations of spatially resolved stars in dwarf galaxies also suggest that while ellipticals and spheroidals have one or more discrete episodes of star formation, gas-rich dwarf irregulars experience quasi-continuous star-formation through-out their history \citep{2003dhst.symp..128T}. The quiescent counterparts of BCD galaxies (QBCD) were detected in SDSS \citep{2008ApJ...685..194S} with a population ratio between QBCD and BCD as 30:1. If all types of dwarf galaxy population is taken in to account, there is about one BCD galaxy for 90 dwarf galaxies, implying that BCD galaxies are indeed rare. \citet{1998astro.ph..5042L} found that quiescent dwarfs and low surface brightness galaxies sustain a low rate of star formation continuously even in their quiescent phases. The colour-magnitude diagram also suggests that the star formation histories of BCD and normal dwarf irregular galaxies are not significantly different except for the current phase of starbursts in BCD galaxies \citep{2010AdAst2010E...3C}. Mrk~22 appears to be witnessing starburst triggered by recent tidal interaction. While other dwarf irregular galaxies in the region may be sustaining star formation through other internal mechanisms such as stochastic self propagating star formation, after an initial burst in the past.  

\subsection{Clues from radio continuum}

We could not associate the radio continuum detected within the optical extent of Mrk~22 to any prominent H{\small{II}} region in the galaxy. If this radio source of $0.9\pm0.3$ mJy flux is somehow related to Mrk~22, the radio luminosity will be $6\pm3 \times$ 10$^{19}$ W Hz$^{-1}$. Radio emission in star forming dwarf galaxies has been studied \citep{1984A&A...141..241K,1986sfdg.conf..281B,1986A&A...161..155K,1991A&A...246..323K,1993ApJ...410..626D,
2003ApJ...597..274Y,2005A&A...436..837H,2011ApJ...728..124R}. These studies indicate that the radio emission in dwarf galaxies is often confined to only some star forming regions and the spectral index is relatively flatter compared to normal spiral galaxies. The WR star forming dwarf galaxies are in general found to be radio deficient \citep{2016MNRAS.462...92J}. It is expected that in very young starbursts such as in WR galaxies, supernovae explosions and hence the relativistic electrons have not been yet produced or have not diffused out over large spatial scales in the galaxy. In very young starbursts, some contribution to total radio emission from emission from young ($\textless$ 0.01 Myr) supernovae remnants can also be expected. The radio power of young extragalactic supernovae remnants detected in the Sedov phase \citep[cf.,][]{1959sdmm.book.....S,1986A&A...166..257B,1998ApJS..117...89G,1999ApJS..120..247G,2004A&A...427..525B} is typically 10$^{16}$ to 10$^{21}$ W Hz$^{-1}$ with a fairly similar luminosity function among galaxies with different SFRs \citep{2009ApJ...703..370C}. The detected radio source in Mrk~22 could be associated with a supernova remnant in some highly evolved and optically faint H{\small{II}} regions in the galaxy. However, there is also a possibility that this is a background radio source not associated with Mrk~22.

The bright star forming region is expected to have high radio flux of $\sim$ 3 mJy, but, is not detected in radio. The estimated age of the bright region as $\sim$ 4 Myr implies that this region is perhaps too young to have supernovae events. This radio non-detection is consistent with the detection of the WR features in this region as the WR phase in the most massive stars appears after 2 to 5 Myr from their birth before the end of this phase through supernova explosions in a very short time of $\textless$ 0.5 Myr \citep{2005A&A...429..581M}. There appears a genuine radio deficiency in Mrk~22 due to lack of supernovae explosions in a young starburst. A lack of large-scale diffuse radio continuum in Mrk~22 also suggests that the recent phase of the starburst in Mrk~22 is triggered after at least a few hundred Myr of quiescence as the galactic synchrotron emission at 1.4 GHz remains detectable for a period of a few hundred Myr after the cessation of massive star formation.

\subsection{Comparison with extreme star-forming galaxies at higher redhsifts}

The BCD galaxies are often considered as local analogs of extreme star-bursting galaxies such as green-pea (GP) galaxies and Lyman-alpha emitters at higher redshifts. Among these extreme systems, the GP galaxies in the redshift range z = 0.11 - 0.35 are the nearest and well studied objects. The GP galaxies are characterized as low mass (M* $\textless$ 10$^{10.5}$ M$_{\odot}$), compact ($\textless$ 5 kpc) star-bursting galaxies with very high [O{\small{III}}] $\lambda$5007 equivalent widths ($\sim$ 1000 \AA). The GPs also exhibit very high [O{\small{III}}] $\lambda$5007/[O{\small{II}}] $\lambda$3726 ratio, on average 10 times higher than the star forming galaxies at lower redshifts \citep{2013ApJ...766...91J}. The [O{\small{III}}] $\lambda$5007/[O{\small{II}}] $\lambda$3726 ratio in nearby galaxies is typically $\textless$ 1 while GP galaxies have this ratio up to 10. Such extreme galaxies are considered as the best candidates for Lyman continuum leakers and ionizing the IGM \citep{2010ApJ...708L..69B,2011MNRAS.412..411Y,2014ApJ...791L..19J,2014MNRAS.442..900N,2015ApJ...809...19H}. The [O{\small{III}}] $\lambda$5007/[O{\small{II}}] $\lambda$3726 ratios for the two star forming regions in Mrk~22 are estimated as 6.16$\pm$2.80 and 3.54$\pm$0.54, respectively. These values are consistent with the finding of \citet{2014ApJ...786..155N} who showed that in optically thin nebulae with optical depth $\textless$ 4, the [O{\small{III}}] $\lambda$5007/[O{\small{II}}] $\lambda$3726 ratio can be higher than 3. This makes physical conditions in the star forming regions in Mrk~22 similar to those in the extreme star forming galaxies. \citet{2013ApJ...769....3N} and \citet{2013koa..prop..229N} indicated that these extreme conditions can be produced in several ways such as density bounded H{\small{II}} regions, very high ionization parameters due to presence of hard ionizing sources, and low metallicity. The presence of WR stars in Mrk~22 and in some GP galaxies indicates high ionization parameter in these galaxies. The R$_{23}$ index for the two regions in Mrk~22 is estimated as 9.12$\pm$1.05 (bright) and 5.75$\pm$0.79 (faint) consistent with the steep linear relation between R$_{23}$ and [O{\small{III}}] $\lambda$5007/[O{\small{II}}] $\lambda$3726 line ratio \citep{2013ApJ...769....3N}. Moreover, the H$\alpha$/H$\beta$ ratio in Mrk~22 is 3.80$\pm$0.49 implying a low extinction E(B-V) = 0.29$\pm$0.04 in Mrk~22, which is also similar to typical value seen in GPs. 

\begin{figure}
\centering
\includegraphics[width=6.5cm,height=8.5cm,angle=270]{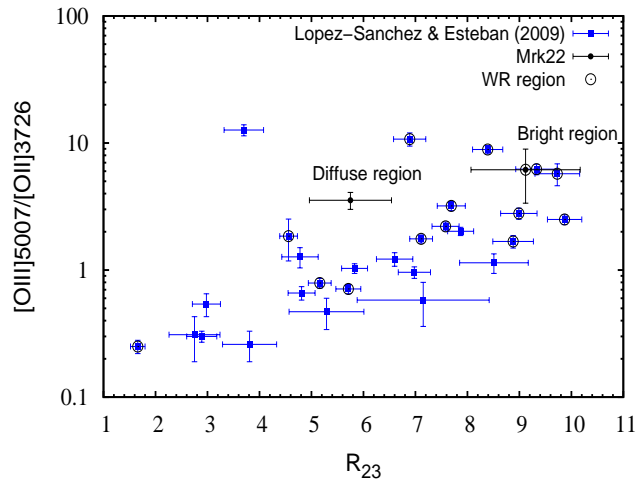}
\caption{Plot of the emission line flux ratio [O{\small{III}}] $\lambda$5007/[O{\small{II}}] $\lambda$3726 against the R$_{23}$ index for two star forming regions (solid circle) in Mrk~22 with other data points (solid square) from \citet{2009A&A...508..615L} associated with several star forming regions in a sample of nearby WR galaxies.}
\label{fig:13}
\end{figure}

It is worth mentioning that EW([{\small{OIII}}] $\lambda$5007) for the brightest region in Mrk~22 is 1620$\pm$452 \AA, which is very high and comparable to those in GP galaxies. However, when comparing EW of GP galaxies with some local analogs, it is important to consider the aperture sizes. The SDSS observations were carried out using 3\arcsec fibre, which contains almost entire GP galaxy at their typical redshifts while a similar size fibre or slit will contain $\textless$ 0.5 kpc region in Mrk~22 or other nearby ($\textless$ 30 Mpc) galaxies. Since EW is an emission line parameter relative to the underlying continuum, GPs spectrum contain almost entire galaxy continuum while the continuum in the spectrum of local analogs will be significantly reduced. Therefore, the local galaxies may show apparently similar or higher EW compared to GPs. Although, the SFR in Mrk~22 is two orders of magnitude lower compared to that in GP galaxies, the ionization and other physical conditions in Mrk~22 are strikingly similar to those in GPs and other extreme star-bursting galaxies at higher redshifts. In order to check occurrences of such extreme star forming regions in nearby galaxies, we plotted [O{\small{III}}] $\lambda$5007/[O{\small{II}}] $\lambda$3726 ratio against R$_{23}$ index in Fig.~\ref{fig:13} for a sample of star forming regions in nearby galaxies with WR features compiled by \citet{2009A&A...508..615L}. This figure indicates that several star forming regions in nearby WR galaxies have ionization conditions similar to the extreme star-bursting galaxies. The above analysis indicates that the physical and ionization conditions responsible for producing extreme ionization in GPs are perhaps not too different from what is seen in some local compact star-bursting galaxies. 

\section{summary and conclusions}  

A multi-wavelength study of the BCD galaxy Mrk~22 using optical long-slit spectroscopic, radio interferometric \HI 21 cm-line, and radio continuum observations was presented here. A summary and important conclusions are as follows:

\begin{itemize}

\item {Mrk~22 has two prominent star forming regions with a projected separation of $\sim0.6$ kpc. The age of the bright region with Wolf-Rayet emission lines is estimated as $\sim$ 4 Myr while the age of the faint region without a detectable Wolf-Rayet emission is estimated as $\sim$ 10 Myr. No diffuse non-thermal radio continuum emission at 1.4 GHz is detected from Mrk~22 indicating a radio-deficiency very likely due to lack of supernovae in the young star-bursting regions. Mrk~22 also appears to have the present starburst after at least a few hundred Myr of quiescence.}

\item {The gas-phase metallicities of bright and faint regions are $\sim$ 8 and $\sim$ 7.5 respectively. The metallicity difference of $\sim$ 0.5 dex between two star forming regions at close separation is significant. The electron temperatures from [O{\small{III}}] lines are estimated as $\sim14000$ K and $\sim19000$ K for the low metallicity (bright) region and high metallicity (faint) region respectively. A lower electron temperature inferred for the high metallicity region is consistent with the general trend seen in the Galactic H{\small{II}} regions. The large metallicity difference within the galaxy can be attributed to a recent merger.}

\item {The \HI morphology and velocity field indicate that Mrk~22 is a disturbed system with \HI plumes like features extending well outside the optical disk. It is very likely a merger system with another dwarf or low surface brightness galaxy. The total \HI mass of the system is estimated as $\sim10^{8}$~M$_{\odot}$.}

\item {Mrk~22 resides in a small group comprising of mainly dwarf irregular galaxies without any massive galaxy. All dwarf irregular galaxies in the group appear blue indicative of ongoing star formation. It is very likely that the galaxies in this group are sustaining star formation through some internal mechanism such as stochastic self propagation star formation, after a burst in the past. The tidal interactions appear as the most viable mechanism for triggering star-burst in these galaxies.}

\item {Mrk~22 shows a high ratio of [O{\small{III}}] $\lambda$5007/[O{\small{II}}] $\lambda$3726 and high [O{\small{III}}]$\lambda$5007 equivalent width. The ionization conditions in Mrk~22 are surprisingly similar to those detected in green pea and extreme star-bursting galaxies at intermediate/high redshifts.}

\end{itemize}
  

\section*{Acknowledgements}
We thank the referee for valuable comments which improved the clarity of the paper. The Image Reduction and Analysis Facility {\small (IRAF)} is distributed by the National Optical Astronomical Observatories (NOAO), which are operated by the AURA Inc., under cooperative agreement with the National Science Foundation (NSF). The FIRST sky survey was conducted by the National Radio Astronomical Observatory (NRAO) using the VLA. The NRAO is a facility of the National Science Foundation operated under cooperative agreement by Associated Universities, Inc. This research has made use of the NASA/IPAC Extragalactic Data base (NED) and the Smithsonian Astrophysical Observatory (SAO)/NASA Astrophysics Data System (ADS) operated by the SAO under a NASA grant. We thank the staff of the HCT and the GMRT who made the observations possible. The HCT is operated by Indian Institute of Astrophysics (IIA) through dedicated satellite communication from the Centre for Research \& Education in Science \& Technology (CREST), IIA, Hosakote, Bangalore, India. The GMRT is operated by the National Centre for Radio Astrophysics (NCRA), Pune, India.

\bibliography{references}

\bsp	
\label{lastpage}
\end {document}